Roles and Needs of Laboratory Astrophysics in NASA's Space and Earth Science Mission

Submitted by the

American Astronomical Society Working Group on Laboratory Astrophysics
http://www.aas.org/labastro/

Nancy Brickhouse - Harvard-Smithsonian Center for Astrophysics
nbrickhouse@cfa.harvard.edu, 617-495-7438

John Cowan - University of Oklahoma
cowan@nhn.ou.edu, 405-325-3961

Paul Drake - University of Michigan
rpdrake@umich.edu, 734-763-4072

Steven Federman - University of Toledo
steven.federman@utoledo.edu, 419-530-2652

Gary Ferland - University of Kentucky
gary@pa.uky.edu, 859-257-879

Adam Frank - University of Rochester
afrank@pas.rochester.edu, 585-275-1717

Wick Haxton - University of Washington
haxton@u.washington.edu, 206-685-2397

Eric Herbst - Ohio State University
herbst@mps.ohio-state.edu, 614-292-6951

Keith Olive - University of Minnesota
olive@physics.umn.edu, 612-624-7375

Farid Salama - NASA/Ames Research Center
Farid.Salama@nasa.gov, 650-604-3384

Daniel Wolf Savin* - Columbia University
savin@astro.columbia.edu, 1-212-854-4124,

Lucy Ziurys – University of Arizona
lziurys@as.arizona.edu, 520-621-6525

[*] Editor

# 1. Introduction

Laboratory astrophysics and complementary theoretical calculations are the foundations of astronomy and astrophysics and will remain so into the foreseeable future. The mission enabling impact of laboratory astrophysics ranges from the scientific conception stage for airborne and space-based observatories, all the way through to the scientific return of these missions. It is our understanding of the under-lying physical processes and the measurements of critical physical parameters that allows us to address fundamental questions in astronomy and astrophysics. In this regard, laboratory astrophysics is much like detector and instrument development at NASA. These efforts are necessary for the success of astronomical research being funded by NASA. Without concomitant efforts in all three directions (observational facilities, detector/instrument development, and laboratory astrophysics) the future progress of astronomy and astrophysics is imperiled. In addition, new developments in experimental technologies have allowed laboratory studies to take on a new role as some questions which previously could only be studied theoretically can now be addressed directly in the lab. With this in mind we, the members of the AAS Working Group on Laboratory Astrophysics (WGLA), have prepared this White Paper on the laboratory astrophysics infrastructure needed to maximize the scientific return from NASA's space and Earth sciences program.

The field of laboratory astrophysics comprises both theoretical and experimental studies of the underlying physics that produce the observed astrophysical processes. We have identified six areas of physics as relevant to astronomy and astrophysics. Astronomy is primarily an observational science detecting photons generated by atomic, molecular, and solid matter physics. Our understanding of the universe also relies on knowledge of the evolution of matter (nuclear and particle physics) and of the dynamical processes shaping it (plasma physics). Hence, our quest to understand the cosmos rests firmly on scientific knowledge in atomic, molecular, solid matter, nuclear, particle, and plasma physics. Chemistry is implicitly included here as part of molecular physics. Additionally, it is worth noting that there is not always a 1-to-1 correspondence between observational band-passes and the needed laboratory astrophysics. For example, standard UV/visible diagnostics for probing astrophysical environments are redshifted to longer wavelengths in high $z$ objects. Also, models of chemical processes involving photons at one wavelength are used to understand environments at other wavelengths.

This White Paper is organized as follows: In Section 2 we highlight some of the recent notable mission-enabling accomplishments of laboratory astrophysics. Section 3 presents a brief historical overview of the funding and infrastructure support for laboratory astrophysics. In Section 4 we highlight some of the key issues necessary to ensure a healthy and vital laboratory astrophysics community. Proposed actions to be taken and estimated levels of support needed are provided in Section 5. In Section 6 we give a brief summary of the current state of the field and in Section 7 we list our recommendations to ensure a vibrant laboratory astrophysics community in the coming decade. A more detailed discussion of the expected scientific impact of laboratory astrophysics in the next decade is given in Appendices I-V.

The findings presented here are based largely on material prepared by the WGLA for the *National Research Council Astronomy and Astrophysics Decadal Survey* (Astro2010). The body of this white paper is drawn from the Position Paper submitted to the *Subcommittee on the State* *of the Profession*, in specific to the *Facilities, Funding, and Programs (FFP) Infrastructure Study Group*. The reports presented in the Appendices were prepared for the *Subcommittee on Science* and the five associated *Science Frontier Panels: Cosmology and Fundamental Physics (CFP), Galactic Neighborhood (GAN), Galaxies across Cosmic Time (GCT), Planetary Systems and Star Formation (PSF), and Stars and Stellar Evolution (SSE).*

This report also builds on the findings from the 2006 NASA Laboratory Astrophysics Workshop which was charged by the NASA Science Mission Directorate (SMD) Universe Working Group (UWG) to draw up a White Paper (NASA Laboratory Astrophysics Workshop White Paper 2006)

1. "addressing the major science goals as defined in the 2006 NASA Strategic Plan and then providing details on the critical laboratory astrophysics data requirements that will have to be met, if the desired science results are actually to be achieved,

2. reporting of recent significant astronomical results where the input from laboratory astrophysics was of critical importance, and

3. discussing in detail the specific laboratory astrophysics efforts that will need to be undertaken in direct support of missions and programs that are on the near horizon, specifically Herschel, SOFIA, JWST, Hubble servicing, and ALMA (the latter of primary concern to NSF)."

## 2. Recent Notable Mission-Enabling Successes of Laboratory Astrophysics

Astrophysical discoveries are propelled forward in part by experimental and theoretical advances in laboratory astrophysics. Presented below are selected examples of significant astrophysical results that arose from recent laboratory and theoretical efforts in this field. This incomplete selection is taken largely from the White Paper from the *Proceedings of the 2006 NASA Laboratory Astrophysics Workshop*.

- New laboratory data, specifically experimental transition probabilities and radiative lifetimes have played a critical role in recent astronomical abundance studies (e.g., Lawler et al. 2006). These new experimental data have led to much more accurate and improved abundances, particularly for the rare earth elements, for the halo stars in the Galaxy, among the oldest stars in the Universe. As a result of these studies, and the increasingly more precise abundance determinations, we have gained a much deeper understanding of the nuclear processes involved in the formation of the heavy elements. Thus, for example, abundance determinations for old metal-poor halo stars using the Hubble Space Telescope and ground-based observations suggest that two different rapid neutron-capture processes may exist for nucleosynthesis beyond the iron peak (Cowan et al. 2005). This insight is the result of new laboratory oscillator strengths for high Z elements (e.g., Ivarsson et al. 2003). These new data have also allowed for improvements in radioactive dating using Th/Eu ratios and are yielding reasonable cosmochronometric age estimates for halo stars (Sneden et al. 2008).

- In the well-studied AGN NGC 3783, the warm absorber density and location have been found, respectively, to be smaller and closer to the central black hole than expected (Krongold et al. 2005). This discovery is a major success of theoretical atomic physics, which identified the numerous unknown absorption lines in the high resolution Chandra and XMM-Newton spectra of warm absorbers. These lines were shown to be inner-shell absorption transitions for the low charge states of Fe and were also used as a powerful new plasma diagnostic (Behar et al. 2001).

- Measuring the atmospheric temperature of the extra-solar planets TrES-1 (Charbonneau et al. 2005) and HD 209458b (Deming et al. 2005) with the Spitzer Space Telescope is partly a success of stellar atmosphere spectral synthesis and laboratory studies of molecular opacities. Transit searches for extra-solar planets also use synthetic stellar spectra as templates against which to correlate measured radial velocities (Konacki et al. 2003), a method which is at the core of NASA's Kepler Mission. For these and other cases, the increasing completeness of the calculated line lists and opacities (Bautista 2004) has been a critical factor.

- Using the Kuiper Airborne Observatory (KAO), important molecules have been detected in the interstellar medium (ISM) whose emission had been inaccessible by ground-based astronomy. Of particular note are $H_2D^+$, a cornerstone species in the ion-molecule theory of interstellar chemistry, HCl, a fundamental hydride, and $H_3O^+$, a direct tracer of the water abundance (e.g., Zmuidzinas et al. 1995; Timmerman et al. 1996), as well as the pure rotational lines of $NH_3$, OH, and CH (e.g., Stacey et al. 1987). The KAO has also been invaluable in detecting ro-vibrational transitions of heavier species such as $C_3$ in molecular clouds (e.g., Giesen et al. 2001). Studies of these species have led to breakthroughs in our understanding of the molecular component of the interstellar medium. These discoveries were only made possible by preceding high resolution, laboratory spectroscopy (e.g., Brown et al. 1993; Harrison et al. 2006). Such laboratory measurements are needed for the sensitive spectral-line surveys proposed for Herschel.

- High rotational lines of CO (up to J=45) have been discovered in the Orion-KL nebula using the KAO and the Infrared Space Observatory (Gonzalez-Alfonso et al. 2002). These spectral lines gave the first early glimpses of the process of high mass star formation with associated shocks and high velocity outflows (e.g., Hollenbach et al. 1995; Ceccarelli et al. 1996). These studies would not have been possible without previous high resolution laboratory spectroscopic work; the understanding of the dynamics in this environment rests on past investigations of collisional excitation. This and the preceding example reveal the exciting results anticipated at sub-millimeter wavelengths with SOFIA and Herschel and from new approaches to molecular synthesis on grains in shock-heated regions (Madzunkov et al. 2006).

- FUSE observed UV absorption of $H_2$ and HD in the ISM from a range of excited rotational levels (Snow et al. 2000, 2008; Ferlet et al. 2000), providing new insights into the physics and chemistry of translucent clouds and a tool to evaluate the ISM deuterium abundance. These observations and their interpretations were made possible by laboratory studies of the electronic transitions and processes relevant to the formation of $H_2$.

- Aromatic infrared bands (AIBs), commonly dubbed "PAH bands", are now routinely observed with the Spitzer Space Telescope and used to probe dust in extragalactic environments (Hogg et al. 2005; Wu et al. 2006) and to measure redshift (Yan et al. 2005; Weedman et al. 2006). Such emission measured in a recent Spitzer survey of elliptical galaxies indicates that the carriers, attributed to polycyclic aromatic hydrocarbons (PAHs) and their derivatives, are a probe of recent merger activity in these galaxies (Kaneda et al. 2005). It certainly looks like the AIB emission is correlated with star formation (Lutz et al. 2005; Wu et al. 2005). This progress has only been possible thanks to laboratory emission and absorption spectra of PAH-type species, including ions. Furthermore, the signature of individual PAH-like molecules can now be sought in astronomical spectra at UV and visible wavelengths, making it possible, for the first time, to detect specific aromatic compounds in space.

- The interpretation of stellar distances using period-luminosity relations for Cephied variables, which involves a very wide range of instruments including for example Spitzer (Freedman et al., 2008), is grounded in detailed opacity calculations by codes such as OPAL. These in turn are grounded upon detailed opacity measurements in plasma systems (Springer, et al. 1997).

- The analysis of the first cometary and interstellar sample returned from the Stardust mission will provide key information on grain formation and processing in space. Observations with ISO and Spitzer have revealed the presence of specific minerals (e.g., crystalline silicates) in a variety of Galactic environments, including outflows from evolved stars and protoplanetary disks (Waelkens et al. 1996; Waters et al. 1996). This work is critical to understanding dust processing during its lifetime, and is only possible thanks to laboratory measurements of the infrared spectra for candidate grain materials (Begemann et al. 1994; Jäger et al. 2003).

- Observations from Chandra of emissions from the Puppis A SNR, in which a shock wave is destroying a clump of material, have been directly interpreted (Hwang et al. 2005) using results from well-scaled laboratory plasma experiments of the same phenomenon (Klein et al. 2003).

- Analysis and interpretation of HST observations of heterogeneous protostelar jet structures were carried out in light of pulsed power laboratory investigations (Lebedev et al 2005, Ciardi et al. 2009). The plasma experiments showed that early onset, non-destructive instabilities would create collimated clumps with dispersion in velocity. HST observations showing heterogeneous or "clumpy" jets (Hartigan & Morse 2007) were then intereped in light of these MHD radiative jet experiments. Astrophysical simulations based on these experiments confirmed the applicability of heterogeneous jets to key observational properties of real YSO outflows (Yirak et al. 2008)

- Chandra observations of Cas A pulsar have been interpreted by Laming et al. (2006) using magnetic tower experiments of Lebedev et al (2005). In the Laming et al. study the parameters of the Cas A plasma jet have been analyzed using the steroidal field acceleration and collimation mechanisms demonstrated in pulsed power laboratory experiments. Laming et al also explored the connections to GRB models.

- Studies of dust and ice provide a clear connection between astronomy within and beyond the solar system (Strazzulla et al. 2005). Planetary surface temperatures are derived from ice measurements in the NIR (Grundy et al. 2002), while planetary atmospheres (Europa, Ganymede, etc.) are explained by laboratory studies of ices (Hansen et al. 2005).

- Recent advances in X-ray astronomy with observatories such as Beppo-SAX, Chandra, XMM_Newton, RXTE, and INTEGRAL have provided a wealth of data on the thermonuclear explosions that accompany accretion onto the surfaces of white dwarfs and neutron stars (X-ray Universe Symposium 2008). The observables depend sensitively on the nuclear physics of reactions for both stable and unstable nuclei, many of which remain poorly measured. Construction will soon begin on the Facility for Rare Isotope Beams, a new accelerator at Michigan State University, which will define for the first time the nuclear physics of neutron-deficient nuclei that participate in the $\alpha$p-process and rapid-proton-capture process powering X-ray bursts (Rare Isotope Science Assessment Committee 2007). Much improved measurements can also be made on stable isotopes using a new generation of low-energy accelerators mounted deep underground, away from cosmic ray backgrounds. Data from the first such facility, LUNA in the Gran Sasso Laboratory, significantly increased age estimates for the oldest globular cluster stars (Broggini et al. 2006) – resulting in a value that overlaps, within errors, WMAP's age for the universe, 13.7 billion years. The NSF hopes to create a second-generation accelerator facility in the US, within the Deep Underground Science and Engineering Laboratory (DUSEL, http://www.jinaweb.org/duesl).

These examples demonstrate the rich astrophysics enabled by laboratory astrophysics. They are, however, by no means the only notable advances. We are limited here only by the constraints of space.

### 3. Historical Overview

In past decades, much of the laboratory astrophysics work required to move astronomy and astrophysics forward was funded by programs in atomic, molecular, solid matter, plasma, nuclear, and particle physics,. In the past, the needs of astronomy and astrophysics were substantially synergistic with the directions of forefront, fundamental research in these fields. As a result, astronomy and astrophysics benefited from laboratory astrophysics research in these six areas without having to support them at a level anywhere close to that required to meet the actual need.

The last decade, however, has seen the funding reality change drastically. A number of NSF and DOE programs that previously supported laboratory astrophysics research are no longer doing so, particularly in the critically important areas of atomic, molecular, and solid matter physics. The research currently supported by these programs has diverged from the needs of the astronomy and astrophysics community. Atomic physics has moved heavily into cold atoms, Bose-Einstein condensates (BEC), and quantum computation and cryptography (collectively known as photonics) as well as ultrafast lasers. Molecular physics has acquired a biological orientation and solid matter physics has moved over to nano-science**.** From a funding perspective, laboratory astrophysics now lies on the boundary between fields and as a result, its support is now insufficient to keep up with the demands of astronomy and astrophysics.

Laboratory plasma astrophysics has for the most part developed more recently. Some of this work benefits at present from an overlap with the fundamental needs of certain subsets of plasma research. Most ongoing and potential research, however, also lies on the boundary between fields and so suffers accordingly.

University support for laboratory astrophysics has also diminished drastically over the past decade as many faculty members have retired and departments have opted to move in new research directions. Areas related to photonics, ultrafast lasers, biology, environmental, or nano-technology are very much in vogue, as opposed to classic atomic, molecular, and solid matter physics. As a result, few new faculty members have been hired in laboratory astrophysics, thus threatening the future supply of researchers knowledgeable in this field. These faculty members are necessary not just to carry out the needed laboratory astrophysics research but also to train graduate and undergraduate students, i.e., the next generation. Student participation in research is critical for the future vitality of the field. The reduced numbers of faculty members and their associated laboratories has also led to a diminishing of the infrastructure in laboratory astrophysics. Most of the instrumentation associated with lab astrophysics research is not commercially available, and loss of personnel also results in a loss of technical expertise - a commodity that cannot easily be replaced.

**4. Ensuring the Future Mission-Enabling Capabilities of Laboratory Astrophysics**

*Increased and steady support for laboratory astrophysics from NASA is critical*. Current support for laboratory astrophysics comes from a small number of insufficiently funded programs. Robust funding programs are necessary to maintain the core competency of the community and to ensure the development of future generations of laboratory astrophysicists. If the funding for current programs is not increased, significant research capabilities that have required decades to develop will be lost. These research programs cannot be turned off and on at will and if stopped would require a large infusion of financial support and many decades to re-achieve previous capabilities. For instance, a new laboratory can cost up to several million dollars, much more than is needed to support, maintain, and enhance current facilities. The impending lack of sufficient and appropriate laboratory astrophysics groups and facilities will impact the scientific return from future astronomy and astrophysics projects, which typically have budgets that dwarf the level of support needed by laboratory astrophysics.

*Explicit support for laboratory astrophysics by missions* is essential to maximize their astronomical and astrophysical scientific return, by using core competencies to deliver data meeting specific needs of the mission. Current laboratory astrophysics funding is insufficient to produce all the critical data needed to ensure successful scientific return from missions and programs. Mission and project support of laboratory astrophysics through competitively run three-to-four year grants at a level comparable to those grants supporting core competency will make a significant impact on the production of the needed laboratory astrophysics data. The current support arising from one-year grants linked to observing cycles does not reflect the long-term nature of laboratory astrophysics research.

*Faculty development in laboratory astrophysics* is necessary to ensure the health and vitality of laboratory astrophysics on university teaching faculties. We urge that NASA offer awards for the creation of new tenure-track faculty positions within the intellectual disciplines that comprise laboratory astrophysics. This is a particularly important issue as start-up packages for laboratory astrophysics hires can be costly. The aim of these awards should be to integrate research topics in laboratory astrophysics into basic physics, astronomy, chemistry, electrical engineering, geosciences, biology, meteorology, computer science, and applied mathematics programs, and to develop laboratory astrophysics programs capable of training the next generation of leaders in this field.

*Establish fellowships and prizes in the area of laboratory astrophysics.* Recognition for research in laboratory astrophysics should be given greater visibility in the community. The creation by NASA of graduate and post-doctoral fellowships in laboratory astrophysics will help to train the next generation of researchers in this field. The establishment of sanctioned awards, analogous to those given in other research areas, would serve to raise the profile of laboratory astrophysics. Possible societies to consider for the creation of such awards include the American Astronomical Society (AAS), the American Chemical Society (ACS), the American Physical Society (APS), or similar professional societies.

*Strong Instrumentation, Technology, and Facilities Development Programs* in laboratory astrophysics are needed to support the development, construction, and maintenance of state-of-the-art laboratory astrophysics instrumentation and facilities. Such programs exist for detector and instrument development for missions. No similar programs dedicated to laboratory astrophysics currently exist. Such programs are vital for ensuring not just that the capabilities of the laboratory astrophysics community remain current with present astronomy and astrophysics needs but that they also prepare for planned future astronomy and astrophysics needs. The time scale for developing new laboratories and technical capabilities is comparable to that for new missions. Correspondingly, support for long term laboratory astrophysics development is critically needed. A range of instrumentation, technology, and development programs at NASA should be developed that would be able to respond to a variety of needs for infrastructure that promote basic research in laboratory astrophysics. The instrument and technology development component should provide funds for the design and construction of state-of-the-art as well as innovative instruments and technologies that will enable new laboratory astrophysics measurements or calculations. The facility development component should provide support for open facilities dedicated to laboratory astrophysics needs.

*Provide adequate funding for databases.* Critically evaluated data are needed by those analyzing astronomical measurements and modeling the associated environments. True understanding is only possible when collections of the highest quality laboratory astrophysics data are utilized. NASA needs to coordinate its efforts with other relevant agencies and departments. Database compilation and the associated, vital critical evaluation, is a skill that is developed over decades in many cases. Long term commitment of funds is essential.

*Maintain a vibrant community of scientists conducting laboratory astrophysics.* The synergy between laboratory work in a university setting and in NASA laboratories must be fostered. A model for funding needs to include a continued level of baseline support, involving strengthened R&A programs, with occasional term-limited spikes for the topical needs of new missions. An example of the latter is the two-year grants in support of the Herschel mission. This is the only viable model to maintain core competency in the field, ensure rapid response to ongoing mission and project needs, and provide a healthy balance among energy bands and among disciplines. Finally, there should be coordination among the federal agencies (NASA, NSF, DOE) that fund all these activities.

*The NRC should conduct a study charged with identifying and detailing the specific laboratory astrophysics needed by the astronomy and astrophysics community.* Such a study should build on the finding of this Space Studies Board study and of the Decadal Survey and follow those findings down to a level of detail which is beyond the scope of this study or of the Decadal Survey. Specific items to address include, but are not limited to, the following:

1. Within the six areas of laboratory astrophysics, what are the specific sub-areas needed by the astronomy and astrophysics community?
2. What is the number of university groups working in each distinct sub-area needed to meet current laboratory astrophysics needs in each sub-area and to train the future generations to insure continued capabilities in each sub-area?
3. How to encourage and retain faculty, in terms of ensuring the future supply of laboratory astrophysicists and maintaining/revitalizing infrastructure in the field?
4. How to foster graduate student participation and Ph.D. theses in these areas?
5. How to support the development and maintenance of laboratories and their unique instrumentation for ground-breaking research?
6. How to coordinate the activities of the agencies and departments that benefit from a robust effort in laboratory astrophysics?
7. How to combine interdisciplinary teams and/or centers focused on solving specific complex problems (e.g., NASA's Astrobiology Institute) while continuing to fund programs to support ground-breaking ideas of individual researchers that could potentially revolutionize aspects of astrophysics and increase the scientific return from observatories?

### 5. Proposed Actions to be Taken and Estimated Levels of Support Needed

At NASA, much of the support for laboratory astrophysics has come through programs such as the Astronomy and Physics Research and Analysis program (APRA). This and other programs historically have supported research in atomic, molecular, and solid matter laboratory astrophysics. Over the past four funding competition cycles the APRA program has received ~ 29 laboratory astrophysics proposals each year. Of these typically ~ 9 are selected for funding each year. Including new and continuing grants, the APRA program supports ~ 27 laboratory astrophysics proposals at a total level of approximately $3 million per year. Considering the vast

scope of atoms, ions, molecules, and solids for which data are needed, and that the need for these data ranges from the far infrared to the x-ray regime, supporting only 9 new such projects each year is woefully inadequate.  Additionally, the average award size of  approximately $110,000 per year is barely enough to support a graduate student researcher (which is needed in order to sustain expertise in laboratory astrophysics over time), one month salary for the PI, computers, conferences, and travel.  Perhaps this may be sufficient for a theory group, but for an experimental group such a level is insufficient to cover the additional expenses of upgrading hardware and supporting technical staff.

If the mandate of the APRA and similar programs remains unchanged, then a trebling of the funding level would make a significant positive impact on the vitality for atomic, molecular, and solid matter laboratory astrophysics.  Such an increase would accomplish two goals.  First, it would provide support for the many highly rated proposals that must currently be declined due to insufficient funds.  Second, an increase in funding will initially increase the odds of receiving an award.  This increase in the likelihood of obtaining funding will serve to draw new researchers into laboratory astrophysics, thereby growing our national capabilities in this area and laying the foundation for a vibrant and strong future.

If, however, the mandate of APRA and similar programs were extended to include plasma, nuclear, and particle laboratory astrophysics, then a dramatically larger increase would be needed.  An expansion of the scope of the laboratory astrophysics covered by these programs without a corresponding increase in the level of funding would have a serious detrimental effect on the laboratory astrophysics capabilities of this nation.  While new research in plasma, nuclear, and particle laboratory astrophysics would be initiated, fewer researchers in atomic, molecular, and solid matter laboratory astrophysics would be able to obtain funding.  The only way to offset changes to the mandates of current programs is to make certain that the changes are accompanied by meaningful increases in the funding level available.  For that reason, we propose that adding plasma, nuclear, and particle laboratory astrophysics to programs such as APRA should be accompanied by at least a doubling of the available funding.  This is in addition to the proposed factor of 3 increase discussed in the preceding paragraph.  Taken together, for example, these changes and increases would grow the laboratory astrophysics portion of the APRA program to a level of approximately $18 million per year.

Announcement of opportunities (AOs) and requests for proposals (RFPs) for all future airborne and space-based observatories should include an explicit request for a detailing of the laboratory astrophysics data needed to analyze the astronomical data from these missions in order to maximize the scientific return.   It is no longer appropriate to think of these data needs as someone else's problem.  The responsibility for ensuring the needed data are there when the observatory comes on line should lie with the proposers of the mission.  For this reason, we propose an appropriate fraction for the budget of each mission be allocated to relevant laboratory astrophysics studies.  Considering the hundreds of millions to billions spent on each mission and the few millions that would be allocated to laboratory astrophysics, it is clear that the lever arm is long and the return on the support for laboratory astrophysics research will be greatly magnified.

Faculty development programs are needed to create the infrastructure at universities necessary to train the next generation of laboratory astrophysicists.  We propose the creation of a program similar to the unrelated NSF Division of Atmospheric Sciences (ATM) Faculty Development program which was closed in 2004. That program funded grants of up to $400,000 per year for up to 5 years. We urge the creation of such programs at the NASA at funding levels similar to those of the closed ATM program but adjusted for inflation.

Instrumentation and technology development programs specifically designed for laboratory astrophysics are vital in order for the capabilities of the laboratory astrophysics community to keep pace with the astronomy and astrophysics needs of this nation. Support for a number of grants each year at a level per three year grant of up to several million dollars per grant would go a long way toward filling this vital need. This program could be modeled on the existing programs at NASA. The need for such programs specifically tailored for laboratory astrophysics is highlighted by the fact that none of these existing instrumentation and technology development programs have a history of supporting laboratory astrophysics.

Facilities and databases require long term commitments extending beyond typical competitively run three year research grants. Additionally the support needed for facilities can often exceed that provided by instrumentation and technology development programs. For example, upcoming missions will study the chemistry of the cosmos. This will require laboratory studies of cold molecules using facilities such as electrostatic ion storage rings. No such facilities currently exist in the United States. The construction capital costs for such a facility are estimated at approximately $15 million and the operating costs at approximately $2 million per year. The US should begin to develop the physics and computational facilities required to meet the long-term laboratory astrophysics needs of astronomy and astrophysics.

## 6. Summary

Laboratory astrophysics has added greatly to the scientific return of NASA missions. Measurements conducted before and after the launch of the Copernicus satellite provided the basis for an improved understanding of interstellar gas. This continued in support of the UV observations conducted with the Goddard High Resolution Spectrograph and the Space Telescope Imaging Spectrograph, where further progress was achieved in both interstellar and stellar research. Many advances at x-ray wavelengths were the combined result of laboratory experiments and related theoretical calculations on the one hand and superb data from Chandra on the other. Laboratory astrophysics is also allowing detailed analyses of solid material seen in spectra with the Spitzer Space Telescope. The community who will soon use the Herschel Space Telescope are keenly aware of their laboratory needs to obtain the greatest scientific impact. Future advances are being jeopardized because the current state of affairs in laboratory astrophysics is dire.

Ensuring the vitality and future of astronomy and astrophysics requires strong national support for a healthy and viable laboratory astrophysics program. Over the past 10 years there have been a series of NASA-sponsored workshops dedicated to the assessment of the state of laboratory astrophysics (NASA Laboratory Astrophysics Workshop 1998, 2002, 2006). These workshops have tracked the increasing direness of the situation. The reports and white papers resulting from these workshops have been published and are listed in the references section.

Laboratory astrophysics has reached a point where it is ceasing to be a viable, productive field, just at a time when advances in experimental technology are opening new vistas in its applicability. We urge the Space Studies Board committee studying mission-enabling activities in NASA's space and Earth sciences program to consider the laboratory astrophysics infrastructure needs for astronomy and astrophysics. We also urge the committee to follow the prioritized proposals they make all the way through to the underlying laboratory astrophysics data needed and to recommend sufficient funding to ensure the future mission-enabling capabilities for the long term vitality of laboratory astrophysics and thereby of astronomy and astrophysics as a whole.

Without laboratory astrophysics, the scientific return from current and future NASA missions will diminish significantly. Funding for laboratory astrophysics has a long lever arm and any additional support will have significant impact on missions and facilities. Without laboratory astrophysics the future progress of astronomy and astrophysics will be greatly hindered.

## 7. Recommendations

- **Increased and steady support for laboratory astrophysics at NASA is critical.**

- **Explicit support for laboratory astrophysics by missions is essential to maximize their astronomical and astrophysical scientific return.**

- **Faculty development in laboratory astrophysics is necessary to ensure the health and vitality of laboratory astrophysics on university teaching faculties.**

- **Establish fellowships and prizes in the area of laboratory astrophysics.**

- **Strong Instrumentation, Technology, and Facilities Development Programs in laboratory astrophysics are needed to support the development, construction, and maintenance of the state-of-the-art laboratory astrophysics instrumentation and facilities that are vital to the success of pending and future missions.**

- **Provide adequate funding for critically evaluated databases which are needed by those analyzing astronomical measurements and modeling the associated environments.**

- **Maintain a vibrant community of scientists conducting laboratory astrophysics.**

- **The NRC should conduct a study charged with identifying and detailing the specific laboratory astrophysics needed by the astronomy and astrophysics community.**

# 8. References

Bautista 2004, A&A, 420, 763.
Begemann et al. 1994, ApJ, 423, L71.
Behar et al. 2001, ApJ, 563, 497.
Broggini et al. 2006, Prog. Part. Nucl. Phys., 57, 343.
Brown et al. 1993, ApJ, 414, L125.
Ceccarelli et al. 1996, ApJ, 471, 400.
Charbonneau et al. 2005, ApJ, 626, 523.
Ciardi et al. 2009, ApJL, 691, L147.
Cowan et al. 2005, ApJ, 627, 238.
Deming et al. 2005, Nature, 434, 704.
Ferlet et al. 2000, ApJ, 538, 69.
Freedman et al. 2008, ApJ 679, 71.
Giesen et al. 2001, ApJ, 551, L181.
Gonzalez-Alfonso et al. 2002, A&A, 386, 1074.
Grundy et al. 2002, Icarus, 155, 486.
Hansen et al. 2005, Icarus, 176, 305.
Harrison et al. 2006, ApJ, 637, 1143.
Hartigan & Morse 2007, ApJ, 660, 426.
Hogg et al. 2005, ApJ, 624, 162.
Hollenbach et al. 1995, ASP Conf. Ser., 73, 243.
Hwang et al. 2005, ApJ, 635, 355.
Jäger et al. 2003, J. Quant. Spectr. Rad. Transf., 79, 765.
Ivarsson et al. 2003, A&A, 409, 1141.
Kaneda et al. 2005, ApJ, 632, L83.
Klein et al. 2003, ApJ, 583, 245.
Konacki et al. 2003, Nature, 421, 507.
Krongold et al. 2005, ApJ, 622, 842.
Laming et al. 2006, ApJ, 644, 260.
Lawler et al. 2006, ApJS, 162, 227.
Lebedev et al. 2005, MNRAS, 361, 97.
Lutz et al. 2005, ApJ, 625, L83.
Madzunkov et al. 2006, Phys. Rev., A 73, 020901(R).
NASA Laboratory Astrophysics Workshop White Paper 2006, S. Federman, N. Brickhouse, V. Kwong, F. Salama, D. Savin, P. Stancil, J. Weingartner, L. Ziurys, in NASA Laboratory Astrophysics Workshop, Ph. Weck, V. Kwong, F. Salama, editors, NASA Conference Proceedings, NASA STI, NASA-Ames Research Center, NASA/CP 214549, 1 (2006); (http://www.physics.unlv.edu/labastro/whitepaper.html).
NASA Laboratory Astrophysics Workshop White Paper 2002, F. Salama, D. Leckrone, J. Mathis, M. McGrath, Miller, R. T. Phillips, W. Sanders, P. Smith, T. Snow, A. Tielens, in NASA Laboratory Astrophysics Workshop, F. Salama editor, NASA Conference Proceedings, NASA STI, NASA-Ames Research Center, NASA/CP 211863, 3 (2002); (http://www.astrochemistry.org/nasalaw.html) .
NASA Laboratory Astrophysics Workshop Report 1998, W. H. Parkinson, K, Kirby, and P. L. Smith, editors , Harvard-Smithsonian Center for Astrophysics, internal document.

Rare Isotope Science Assessment Committee 2007, "Scientific Opportunities with a RARE-ISOTOPE FACILITY in the United States", The National Academies Press.
Sneden et al. 2008, ARA&A, 46, 241.
Snow et al. 2000, ApJ, 538, 65.
Snow et al. 2008, ApJ, 688, 1124.
Springer, et al. 1997, J. Quant. Spectrosc. Radiat. Transf., 58, 927.
Stacey et al. 1987, ApJ, 313, 859.
Strazzulla et al. 2005, Icarus 174, 31.
Timmerman et al. 1996, ApJ, 463, L109.
Waelkens et al. 1996, A&A, 315, L245.
Waters et al. 1996, A&A, 316, L361.
Weedman et al. 2006, ApJ, 638, 613.
Wu et al. 2005, ApJ, 632, L79.
Wu et al. 2006, ApJ, 639, 157.
Yan et al. 2005, ApJ, 628, 604.
Yirak et al. 2009, ApJ, in press.
X-ray Universe Symposium 2008, http://xmm.esac.esa.int/external/xmm_science/workshops/2008symposium/
Zmuidzinas et al. 1995, ASP Conf. Ser., 73, 555.

**Appendix I: New Discoveries in Cosmology and Fundamental Physics through Advances in Laboratory Astrophysics**

## 1. Introduction

As the Cosmology and Fundamental Physics (CFP) panel is fully aware, the next decade will see major advances in our understanding of these areas of research. To quote from their charge, these advances will occur in studies of "the early universe, the microwave background, the reionization and galaxy formation up to virialization of protogalaxies, large scale structure, the intergalactic medium, the determination of cosmological parameters, dark matter, dark energy, tests of gravity, astronomically determined physical constants, and high energy physics using astronomical messengers."

Central to the progress in these areas are the corresponding advances in laboratory astrophysics which are required for fully realizing the CFP scientific opportunities within the decade 2010-2020. Laboratory astrophysics comprises both theoretical and experimental studies of the underlying physics which produce the observed astrophysical processes. The 5 areas of laboratory astrophysics which we have identified as relevant to the CFP panel are atomic, molecular, plasma, nuclear, and particle physics.

Here, Section 2 describes some of the new scientific opportunities and compelling scientific themes which will be enabled by advances in laboratory astrophysics. In Section 3, we provide the scientific context for these opportunities. Section 4 briefly discusses some of the experimental and theoretical advances in laboratory astrophysics required to realize the CFP scientific opportunities of the next decade. As requested in the Call for White Papers, Section 5 presents four central questions and one area with unusual discovery potential. Lastly, we give a short postlude in Section 6.

## 2. New scientific opportunities and compelling scientific themes

Some attempts to unify gravity with the strong and electroweak forces suggests the possibility for temporal and spatial variations of the fundamental "constants" such as the fine-structure constant $\alpha = e^2/\hbar c$ and the electron/proton mass ratio $\mu = m_e/m_p$ (Uzan 2003). Astrophysical research is uniquely situated to address the question of whether these constants are so or if they vary over the lifetime of the universe.

Measuring cosmological parameters has reached amazing levels of precision. Data from the angular power spectrum of the CMB can be used to determine several key cosmological parameters including relative densities of baryonic matter, dark matter, and dark energy as well as test models of inflation (Wong et al. 2008). Observations can also constrain the specific entropy of the universe (Sunyaev & Chluba 2008), the primordial helium abundance (Peimbert et al. 2008), and the CMB monopole temperature (Hamann & Wang. 2008). Studies in these areas lie on the cutting edge of cosmology.

Cosmologists want to probe the structure and reionization of the universe during the dark ages, the period between recombination and the formation of the first stars (Pritchard & Loeb 2008). Such observations are expected to begin in the next decade.

There is now a multitude of evidence for the existence of non-baryonic dark matter. In addition to the above mentioned CMB measurements, there is a host of astronomical observations including the measurement of galactic rotation curves, X-ray emission from clusters of galaxies, and gravitational lensing. All observations are consistent with dark matter comprising roughly 20% of the total energy density in the Universe. Numerous laboratory experiments are currently actively searching for the direct detection of dark matter. A positive detection from these experiments coupled with future data from the Large Hadron Collider (LHC) at CERN may determine the identity of the dark matter in addition to firmly establishing its existence.

Studies of the IGM and the Lyman α forest can be used to constrain the spectral shape and history of the metagalactic radiation field, the chemical evolution of the universe, and the initial mass function of the earliest generation of stars.

### 3. Scientific context

In the standard model of electroweak interactions, fermion masses as well as the weak gauge boson masses are generated through the Higgs mechanism. The standard model contains a fundamental scalar field, the Higgs boson, with self-interactions which allow for the dynamical generation of a non-zero vacuum expectation value. As a consequence, all particles, interacting with the Higgs boson, acquire mass. In an analogous manner, unified theories such as string theory contain additional scalar fields for which there is no fixed background value. If these fields couple to electric and magnetic fields, their background value, which may evolve in an expanding universe, determine the gauge coupling constants, such as the fine structure constant. In these unified theories, variations of the gauge couplings may lead to variations of Yukawa couplings (the couplings of the Higgs bosons to fermions) as well as the variation of the scale at which strong interactions become non-perturbative $\Lambda_{QCD}$. As a result, in addition to variations in α, variations in the ratios of particle masses such as $m_e/m_p$ may occur.

Quasar absorption spectral observations have been used to probe for variations in fundamental constants (King et al. 2008, Murphy et al. 2008). Variations in α can be determined from detailed atomic and molecular absorption features. One particularly sensitive test is the many-multiplet method based on the relativistic corrections to atomic transitions using several lines from various elemental species and allows for sensitivities which approach the level of $10^{-6}$. Similarly, measurements for the variability of μ rely on Lyman and Werner transitions of the $H_2$ molecule.

While the standard model of cosmology can be simply formulated in the context of general relativity, a precise description relies on several key parameters which include the relative densities of baryonic and dark matter, as well, as the dark energy, the rate of expansion of the universe characterized by the Hubble parameter, the shape of the angular power spectrum, the optical depth to the last scattering of the CMB, as well as others. *WMAP* has achieved an enormous leap in the accuracy of these parameters and with the launching of *Planck*, the third generation CMB satellite, a new level of precision will be achieved for CMB temperature and polarization anisotropies (Wong et al. 2008). These measurements can be used to constrain various models for inflation.

CMB observations at decimeter wavelengths are planned for the near future. Any detected CMB distortions will provide an additional means to determine some of the key parameters of the universe such as the specific entropy of the universe, the CMB monopole temperature, and the primordial helium abundance (Sunyaev & Chluba 2008).

Plans have been made to probe the dark ages through observations of the redshifted 21-cm hyperfine line in neutral hydrogen (Bowman et al. 2007). The first generation experiments to carry out these observations are currently under construction.

The first indication for dark matter came from the observations of high velocities of galaxies within clusters of galaxies and required an additional source of gravity beyond that which could be accounted for from light-producing galaxies in the cluster. Similarly, the rotation of stars and gas in spiral galaxies also pointed to the notion that galaxies were embedded in a large and massive halo of dark gravitating matter. The presence of hot X-ray emitting gas also requires a significant amount of gravity to prevent the hot gas from flying off into space. Observations of distant galaxies along the line of sight of a large cluster of galaxies show clear signs of the gravitational lensing of the light. Indeed, recent claims to direct evidence for dark matter came from observations of a collision of clusters of galaxies showing lensing which is directly associated with (non-dissipative) dark matter in tact with the two clusters whereas the (dissipative) hot gas has been stripped from the cluster by the collision (Clowe 2006).

The standard model of strong and electroweak interactions has been very well established by precision measurements at LEP and SLAC. At the energies achieved so far, there are precious few discrepancies which can be taken as a sign of new physics beyond the standard model. Of course a key missing constituent of the standard model is the Higgs boson. Its discovery is the primary goal for the LHC which will begin operations this year. Nevertheless, particle physics models beyond the standard model are known to possess particle candidates for the dark matter. For example, axions are particles which appear in models which attempt to solve the so-called strong CP problem associated with the strong nuclear force.

One problem associated with the presence of fundamental scalar particles in the standard model is known as the hierarchy problem. Put simply, in the context of a unified theory with physics at a very high energy scale, it is very difficult to maintain the scale of the breakdown of the electroweak theory to energies as low as 100 GeV. An elegant solution to this problem is known as supersymmetry which is a theory relating particles which differ in their spin statistics by a ½ integer (in units of ℏ). Each known particle in the standard model is associated with a superpartner, yet to be discovered. The discovery of superpartners is also one of the main goals for the LHC. The unification of running gauge couplings at high energy in unified theories also requires an extension of the standard model similar to the supersymmetric extension. The lightest of all the new supersymmetric particles is expected to be stable as its decay is protected by the conservation of a new quantum number in much the same way that the proton is protected by the conservation of baryon number or the electron by the conservation of electric charge. This new massive particle is an ideal candidate for dark matter.

The planned dramatic increase in collecting area for the next generation of ground-based observatories, going from 10 m to 30 m telescopes, is expected to revolutionize observational and theoretical studies of the IGM and the Lyman α forest. Such studies will dramatically improve our ability to constrain the spectral shape and history of the metagalactic radiation field, the chemical evolution of the universe, and the initial mass function of the earliest generation of stars.

### 4. Required advances in Laboratory Astrophysics

Advances particularly in the areas of atomic, molecular, plasma, nuclear, and particle physics will be required for the scientific opportunities described above. Here we briefly discuss some of the relevant research in each of these 5 areas of laboratory astrophysics. Experimental and theoretical advances are required in all these areas to fully realize the CFP scientific opportunities of the next decade.

#### 4.1. Atomic Physics

Spectroscopic observations of quasar absorption systems to search for variations in the fundamental constants require accurate transition frequencies and their dependence on these constants for many different atomic systems. Strong systematic uncertainties arise from inaccurate laboratory measurements and some systems of particular importance include ions of Ti, Mn, Na, C, and O (Berengut et al. 2004). Atomic clocks and laboratory transition frequency determinations can also be used to constrain variations in the fundamental constants (Lea 2007).

Interpreting observations from *Planck* will require a precise knowledge of the cosmological recombination history. The complete lack of understanding of the profile of the CMB last-scatter surface dominates the error budget for calculating the cosmological power spectrum. Reducing these uncertainties requires high precision calculations for the cosmological recombination process which in turn rely on accurate data for hydrogen and helium recombination and related processes (Wong et al. 2008).

Observations of the cosmological recombination spectrum can in principle be used to measure the specific entropy of the universe, the primordial He abundance, and the CMB monopole temperature. Interpreting these observations will also require accurate atomic data for recombination spectra in hydrogen and helium and particularly for two photon processes (Sunyaev & Chluba 2008).

Interpreting the planned red-shifted 21-cm observations will require a detailed model of the underlying atomic processes affecting the population of the hyperfine levels. These include resonant scattering of Lyman-alpha photons, spin changing collisions of the hyperfine levels, and related processes (Pritchard & Loeb 2008).

Observations of H II regions are used to infer primordial helium abundances to test the standard model for big bang nucleosynthesis (BBN). Reliable recombination rate coefficients are needed for the observed H I and He I lines as well as for ions of O, Ne, Si, and S (Peimbert et al. 2007).

Interpreting spectra of the IGM Lyman α forest is carried out using both single-phase models and cosmological models of the IGM employing semianalytical approximations for hydrodynamical simulations. These various models used different approximations and assumptions. However, one thing they all have in common is the need to calculate the ionization structure of the photoionized IGM. This is typically carried out using plasma codes written specifically for modeling the ionization structure of photoionized gas. One of the most commonly used codes for this purpose is CLOUDY (Ferland et al. 1998). Fundamental to the accuracy of such plasma codes and any inferred astrophysical conclusions is calculating the correct ionization balance. This requires reliable photoionization, radiative recombination, dielectronic recombination, and charge transfer data (Savin 2000). Accurate data are also needed for photoabsorption (photoexcitation), electron impact excitation, and radiative lifetimes to generate accurate model spectra.

### 4.2. Molecular Physics

Quasar observations of molecular absorption lines can be used to constrain temporal and spatial variations in fundamental constants. For this accurate transition frequencies and their dependence on these constants for many different molecular systems are needed. Some of the more important systems include $H_2$ (King et al. 2008) and OH (Kanekar et al. 2005). Molecular clocks and laboratory transition frequency studies can also be used to constrain variations in the fundamental constants (Lea 2007).

### 4.3. Plasma Physics

Plasma physics at high energy density can produce photoionized plasmas that can benchmark models of the IGM and the reionization of the universe. This is accomplished by producing an intense x-ray source that can irradiate a plasma volume containing relevant species, and measuring the ionization balance and other properties that result. Work in this direction has begun (Foord et al. 2006; Wang et al. 2008). Much more will be possible in the coming decade as higher-energy facilities come online.

### 4.4. Nuclear Physics

A critical intersection between cosmology and laboratory nuclear astrophysics comes from the role of neutrino mass. Neutrinos are the one identified component of the dark matter, though laboratory bounds and cosmological observations rule out neutrinos as the principal component. Their effect on cosmological evolution, however, is not neglible and oscillation experiments place a lower bound on the contribution. This is important because cosmology can determine the sum of the neutrino masses, a quantity that cannot be extracted from oscillation experiments, which test only $m^2$ differences. The mass scale, in cosmological analyses, is correlated with the parameters of the dark energy
equation of state and other neutrino-physics inputs, such as the possibility of additional sterile neutrinos. Thus an independent determination of this scale would be significant.

The best laboratory limit on the sums of the neutrino masses comes from tritium beta decay in combination with oscillations, yielding a bound of 6.6 eV. The inadequacy of this limit is apparent from the more model dependent, but more stringent, result from cosmological analyses of about 0.7 eV. A new tritium endpoint experiment is underway (KATRIN) that is expected to improve this laboratory limit by an order of magnitude, thus providing a check on the current cosmological bound. There are also ambitious new double beta decay experiments that probe, somewhat less directly, the neutrino mass scale. These experiments - carried out by groups in both nuclear and particle physics - have much greater reach and could potentially establishing bounds on the order of 0.05 eV, if mounted at the requested one-to-ten-ton scale. It is anticipated that the cosmological bound will also improve rapidly: analyses based on PLANCK plus SDSS and on CMBpol plus SDSS may be able to reach 0.21 and 0.13 eV, respectively. With future data from weak lensing surveys, this bound could move below 0.1 eV.

The concordance between laboratory and cosmological neutrino mass determinations could be of great importance. One can envision a cosmological mass determination and measurement of the currently unknown third mixing angle (the subject of current reactor and long-baseline programs) in the next decade. In total there would be six constraints (two mass differences, three mixing angles, the mass scale) imposed on the double beta decay Majorana mass. A lack of concordance would yield new physics, e.g., constraining the relative CP of the mass eigenstates and/or their Majorana phases.

The detection of neutrinoless double beta decay would have another important implication for cosmology: it is the only practical low-energy tool for establishing lepton number violation and Majorana masses, ingredients in theories that invoke leptogenesis as a mechanism for generating the cosmological baryon number asymmetry.

The first stars are thought to have formed ~ 100 Myr after the Big Bang: dark matter halos of $10^{5-6}$ solar masses at redshift $z \sim 30$ attract enough primordial gas to produce significant baryon densities and temperatures. By $z \sim 20$ sufficient cooling has occurred to produce gravitational instability. The first stars played an important role in the evolution of the early universe. Prior to their formation the universe consisted mostly of neutral hydrogen and helium, and was dark due to the absorption of light by these atoms. The ultraviolet photons from these early stars helped reionize the early universe, ending the "cosmic dark age". These stars thus influenced conditions for early galaxy formation (e.g., the first 1 Gyr), through their radiation, heavy element production, kinetic energy input, and generation of shock waves and electromagnetic fields. This epoch is interesting cosmologically because it is accessible: WMAP and studies of distance quasars by SDSS have provided some information on the universe's reionization history, while VLT, Subaru, and Keck have helped us understand the early evolution of metals.

Nuclear physicists are involved in this interdisciplinary field through their efforts to understand the structure, very rapid evolution, and death of these massive, extremely low metalicity stars - and how this evolution affects the interstellar medium. The work includes the evolution of progenitor models and large-scale numerical simulations to understand the mechanisms for their deaths (e.g., pair instability supernovae vs. collapse to a black hole). They are also very interested in understanding the patterns of the metals that might emerge from the ejecta - a problem that connects observations of metal-poor stars to laboratory measurements of nuclear masses and decays.

Nuclear transitions are possible candidates for measuring the time variation in α. One system of particular interest is $^{229}$Th (Litvinova et al. 2009). Such studies will require highly accurate nuclear model calculations.

### 4.5 Particle Physics

The search for dark matter has become a multi-faceted research enterprise. The search for candidates associated with supersymmetric theories exists on three fronts: accelerator searches, direct detection searches and indirect detection searchs. Accelerator searches are geared towards new particle discoveries. Supersymmetry predicts many new massive particles and the discovery of any of these particles would go a long way in our understanding of the possible nature and identity of dark matter. However, as it is expected that the dark matter is neutral under strong and electromagnetic interactions, it is not likely that the LHC will detect the dark matter; rather this new particle would be expected to escape the detectors and appear as missing energy much like neutrinos.

Similarly, indirect detection experiments look at by products of interactions of dark matter. For example, dark matter annihilation in the galactic halo could provide unique signatures found in the gamma ray, positron, or antiproton backgrounds. Many experiments are currently running looking for these signatures. Most notable perhaps is Fermi, mapping out the gamma ray sky. Another potential signature comes from the scattering and eventual trapping of dark matter particles in the sun. Annihilations in the sun, would produce high energy neutrinos from the sun and provide a unique signature for dark matter. The Amanda and IceCube detectors active at the South Pole are geared for the indirect detection of dark matter.

In contrast to accelerator and indirect detection searches, direct detection experiments are key to the actual discovery of dark matter. These laboratory experiments look for the elastic scattering of dark matter particles with nuclei in low background environments located deep underground. There has been dramatic improvement in the sensitivity of these detectors over the last several years. These detectors are designed to look for weakly interacting massive particles (WIMPS) with masses between 10 GeV and 10 TeV through the nuclear recoil (with energies between 1 and 100 keV) of an elastic scattering event. There are several running experiments with detectors consisting of a wide range of nuclear targets. The best limits on the elastic scattering cross section come from two experiments, CDMS and XENON10. CDMS, currently located in the Soudan mine in Minnesota, uses cryogenic Ge and Si detectors. XENON10 , located at Grand Sasso, uses liquid Xe as its active component. Current limits are at the level of a few times $10^{-8}$ pb. Both experiments have planned upgrades.

Terrestrial experiments search for axions by looking for their conversion to photons in a strong magnetic field (Sikivie 1983). Two high "Q" cavities are currently running. The ADMX experiment at LLNL uses electronic amplifiers with low noise temperature to enhance the conversion signal. Currently this experiment has set limits on the axion mass, excluding masses between 1.9 and 3.3 μeV if the axion constitutes the dark matter. This experiment is being upgraded with the use of SQUID amplifiers. A second experiment CARRACK in Kyoto uses highly excited atoms to detect microwave photons resulting from the axion conversion. This experiment excludes masses around 10 μeV and is being upgraded to a sensitivity capable of probing masses between 2 and 50 μeV.

### 5. Four central questions and one area with unusual discovery potential
#### 5.1 Four central questions
- Are the fundamental "constants" truly constant?
- How accurately can we measure the cosmological parameters?
- What is the nature and identity of the dark matter?
- What is the shape and history of the metagalatic radiation field, the chemical evolution of the universe, and the initial mass function for the first stars?

#### 5.2 One area with unusual discovery potential
- What is the history and structure of the universe during the dark ages?

### 6. Postlude

Laboratory astrophysics and complementary theoretical calculations are the part of the foundation for our understanding of CFP and will remain so for many generations to come. From the level of scientific conception to that of the scientific return, it is our understanding of the underlying processes that allows us to address fundamental questions in these areas. In this regard, laboratory astrophysics is much like detector and instrument development; these efforts are necessary to maximize the scientific return from astronomical observations.

### References
Berengut, J. C., et al. 2004, Phys. Rev. A, 70, 064101.
Bowman, J. D., et al. 2007, Astrophys. J., 661, 1.
Clowe, D., et al. 2006, Astrophys. J., 648, L109
Ferland, G., et al. 1998, Publ. Astron. Soc. Pacific, 110, 761.
Foord, M. E., et al. 2006, J. Quant. Spectrosc. Radiat. Transfer, 99, 712.
Hamann, J., & Wong, Y. Y. Y. 2008, J. Cosmol. Astropart. Phys., 3, 25.
Kanekar, N., et al. 2005, Phys. Rev. Lett., 95, 261301.
King, J. A., et al. 2008, Phys. Rev. Lett., 101, 251304.
Lea, S. N. 2007, Rep. Prog. Phys., 70, 1473.
Murphy, M. T., et al. 2008, Mon. Not. R. Astron. Soc., 384, 2008.
Peimbert, M., et al. 2008, Astrophys. J., 666, 636.
Pritchard, J. R., & Loeb, A. 2008, Phys. Rev. D. 78, 103511.
Savin, D. W. 2000, Astrophys. J., 533, 106
Sikivie, P. 1983, Phys. Rev. Lett., 51, 1415.
Sunyaev, R. A., & Chluba, J. 2008, Am. Soc. Pacific Conf. Ser., 395, 35.
Uzan, J.-P. 2003, Rev. Mod. Phys., 75, 405.
Wang, F. L., et al. 2008, Phys. Plasmas, 15, 0731

**Appendix II: New Discoveries in the Galactic Neighborhood through Advances in Laboraory Astrophysics**

### 1. Introduction

As the Galactic Neighborhood (GAN) panel is fully aware, the next decade will see major advances in our understanding of this area of research. To quote from their charge, these advances will occur in studies of "the galactic neighborhood, including the structure and properties of the Milky Way and nearby galaxies, and their stellar populations and evolution, as well as interstellar media and star clusters".

Central to the progress in these areas are the corresponding advances in laboratory astrophysics that are required for fully realizing the GAN scientific opportunities within the decade 2010-2020. Laboratory astrophysics comprises both theoretical and experimental studies of the underlying physics and chemistry that produces the observed astrophysical processes. The 5 areas of laboratory astrophysics that we have identified as relevant to the GAN panel are atomic, molecular, solid matter, plasma, and nuclear physics.

In this white paper, we describe in Section 2 some of the new scientific opportunities and compelling scientific themes that will be enabled by advances in laboratory astrophysics. In Section 3, we provide the scientific context for these opportunities. Section 4 briefly discusses some of the experimental and theoretical advances in laboratory astrophysics required to realize the GAN scientific opportunities of the next decade. As requested in the Call for White Papers, Section 5 presents four central questions and one area with unusual discovery potential. Lastly, we give a short postlude in Section 6.

### 2. New scientific opportunities and compelling scientific themes

The significant progress in our understanding of the Galactic neighborhood over the past decade has led to fundamental questions that offer scientific opportunities for the next one. Spectroscopic measurements at millimeter, sub-millimeter, infrared, visible, and ultraviolet wavelengths with high signal to noise and analyses based on new precise atomic, molecular, and nuclear data are providing new insights into chemical evolution. Computer simulations in combination with interferometric observations are indicating the need to describe the interstellar medium (ISM) as a very dynamic entity. Observations in the ultraviolet and x-ray portions of the spectrum are revealing the physical conditions in gas within the halo, the connections between the dynamic ISM in the disk and flows in the halo, and the relationship with extra-planar gas in other galaxies. Improved data of high quality are also probing the dynamic phenomena occurring at the Galactic Center and how the magnetic fields there are playing a role. Observations in the millimeter and infrared are probing the evolution of molecules in interstellar clouds into solid dust grains, leading to key information on the evolution of the ISM. Probably the most exciting opportunity lies in the field of astrobiology, where interdisciplinary efforts are seeking answers to fundamental chemical and biological questions.

### 3. Scientific context

The origin of the elements is one of the central problems of nuclear astrophysics. For elements lighter than iron, the combination of improved atomic data and refined models of stellar atmospheres has yielded precise abundances with improved accuracy (e.g., Scott et al. 2009). Approximately half of the elements heavier than iron were synthesized in billion-degree explosions characteristic of core-collapse supernova, in the process of rapid neutron capture. Observations of old, metal-poor stars have found *r*-process distributions remarkably similar to that of our Sun, suggesting that the *r*-process mechanism is a unique event, operating in the first stars as it does now, producing a characteristic nucleosynthetic signature. Recent progress came about as a result of new laboratory measurements on transition probabilities (e.g., Den Hartog et al. 2006; Lawler et al. 2008). Many of the other heavy elements were produced by slow neutron capture in low mass AGB stars and in the He- and C-burning shells of massive stars. Nuclear data acquired at stellar temperatures (Heil et al. 2008) and modeling of stellar interiors (The et al. 2007; Heil et al. 2008) have greatly improved our understanding of the *s*-process. Stellar models are usually developed to reproduce meteoritic abundances (Lodders 2003). The next level of our understanding of chemical evolution will come from a combination of high quality astronomical spectra and more precise atomic and nuclear data that will be needed to interpret the observations.

Dynamical phenomena control many facets of the ISM, from the formation and destruction of interstellar clouds to the onset of star formation to flows in the Galactic halo. Supernova blast waves and winds from massive stars provide the necessary energy input. Numerical modeling has reached a stage where direct comparison with astronomical observations is providing keen insight into the processes taking place (e.g., de Avillez & Breitschwerdt 2004). Observations from radio (Gaensler & Slane 2006) to x-rays (Hwang et al. 2004) are beginning to reveal the dynamics of supernova remnants and of their interactions with the ISM. Ions seen in ultraviolet and x-ray spectra of the Galactic halo (e.g., Ganguly et al. 2005; Williams et al. 2007) are providing key information on the processes creating the distribution of charge states. The Galactic fountain is a popular model describing such phenomena (Shapiro & Field 1976; Bregman 1980). Controversy, however, surrounds the interpretation of the x-ray measurements. Are the measurements really probing the halo, or instead the intergalactic medium? Similar phenomena appear to be present in the halos of edge-on spiral galaxies (e.g., Tüllman et al. 2006). Further analyses involving observations and simulations will provide insight into this facet of galactic evolution.

The Galactic Center harbors a supermassive black hole, but it appears to be underluminous compared to active galactic nuclei. Studies of flaring events at x-ray wavelengths are likely to provide insight into this difference (e.g., Porquet et al. 2008). Radio observations reveal the presence of non-thermal emission associated with the filamentary nature of the magnetic field there. Such measurements also provide evidence for massive star formation (e.g., Yusef-Zadeh et al. 2008). Our understanding of the nearest active galactic nucleus requires improvements in our knowledge of atomic and molecular plasmas in extreme conditions.

Over the past 30 years, space-based and ground-based astronomy have shown that the Universe is highly molecular in nature. The discovery of over 140 different chemical compounds in interstellar gas, with the vast majority organic molecules, and the realization that obscuring dust pervades vast regions of the interstellar medium have revealed the complexity of interstellar chemistry. Protogalaxies and the first stars are predicted to have formed from primordial clouds where $H_2$ and HD controlled the cooling and collapse of these clouds. Subsequent stars and planetary systems are known to form out of the most complex molecular environments deep inside cold gas and dust clouds, often obscured by hundreds of visual magnitudes of extinction; therefore, it is inevitable that interstellar chemistry is intimately connected to the origins of life. An understanding of the molecular component of the Universe requires a two-fold approach. First, the chemical compounds, their abundances, and how they are distributed in astronomical sources need to be determined. Second, molecular formation mechanisms including reaction pathways and dynamics need to be understood. Attaining these goals is crucial in guiding future astronomical observatories designed to provide new insight into galactic evolution.

## 4. Required advances in Laboratory Astrophysics

Advances particularly in the areas of atomic, molecular, solid matter, plasma, and nuclear physics will be required for the scientific opportunities described above. Here we briefly discuss some of the relevant research in each of these 5 areas of laboratory astrophysics. Experimental and theoretical advances are required in all these areas for fully realizing the GAN scientific opportunities of the next decade.

### 4.1. Atomic Physics

Analyzing and modeling spectra begin with identifying the observed lines that may be seen in emission or absorption. This requires accurate and complete wavelengths across the electromagnetic spectrum. The next step toward understanding the properties of an observed cosmic source depends on accurate knowledge of the underlying atomic processes producing the observed lines. Oscillator strengths and transition probabilities are critical to a wide variety of temperature and abundance studies from infrared to X-ray wavelengths. Many existing data for the heavier elements are still notoriously unreliable. The current limitations on the atomic data available for mid-Z elements make it difficult to determine the nature of the *r*-process.

Modeling ionization structure requires accurate data on many processes. Photoionized gas requires reliable low temperature dielectronic recombination (DR) and electron ionized gas high temperature DR. Calculating low temperature DR is theoretically challenging and for some systems laboratory measurements are the only way to produce reliable data. For high temperature DR, few benchmark measurements exist for L-shell and M-shell ions. Density-dependent DR rate coefficients are needed for dense plasmas. For decades astrophysicists have had to rely on theoretical photoionization calculations of varying degrees of sophistication. The development of third generation synchrotron light sources has opened up the possibility of measuring photoionization cross sections for many important ions. Charge exchange (CX) recombination with H and He and CX ionization with $H^+$ and $He^+$ have been shown to be important for many systems, but few modern calculations or laboratory measurements exist at the relevant temperatures. Data are also needed for low charge states of elements such as Se and Kr in order to study nucleosynthesis in stars that are the progenitors of planetary nebulae.

Line ratios, a key diagnostic, involve knowledge of collision strengths and related phenomena. Rate coefficients for electron impact excitation approaching 10% accuracy are necessary for the most important line ratio diagnostics yielding temperature, optical depth, density, and abundance. Proton impact excitation is important because ions in hot post-shock material decouple from radiatively cooling electrons and may remain hot enough to produce line emission through collisional impact. Furthermore, electron impact ionization (EII) data are highly suspect. Recommended data derived from the same scant set of measurements and calculations can differ by factors of 2 to 3. Much of the published experimental data include contributions from an unknown metastable fraction in the ion beams used. The recommended EII data have not undergone any significant revision or laboratory benchmarking since around 1990. Little data also exist for three-body recombination, the time reverse of EII, which is important in high density plasmas.

### 4.2. Molecular Physics

High-resolution laboratory spectroscopy is absolutely essential in establishing the identity and abundances of molecules observed in astronomical data. Given the advancements in detector technologies, laboratory measurements need to have a resolving power higher than the astronomical instruments. This is essential to interpret ubiquitous interstellar spectral features such as the IR emission bands ("PAH bands") and the diffuse interstellar bands (DIBs) that represent a reservoir of cosmic organic material and hold the key to our understanding of the molecular universe (Salama 1999). Furthermore, all the main functional groups known to organic chemists have now been observed in interstellar molecules, suggesting that the origin of life may have begun in the gas phase chemistry of interstellar clouds. Laboratory spectroscopy is crucial in making the link between interstellar molecules and simple biological compounds that could seed life. It is also crucial in making the link between interstellar molecules in the gas phase and dust grains in the solid phase, a key phase in our understanding of the evolution of interstellar clouds and of circumstellar mass loss. For molecular data obtained from astronomical observations to be of practical use, accurate assignments of observed spectral features are essential. The problem here is two-fold. First, the transitions of known molecules need to be assigned in these spectra, including higher energy levels and new isotopic species. Second, the spectra of undiscovered species that promise to serve as important new probes of astronomical sources need to be identified.

The spectroscopic study of interstellar molecules, many of which are complex structures that cannot be produced in large abundance in the laboratory, requires the development and application of state-of-the-art ultra-sensitive spectroscopic instruments. Detecting the possible presence of a species, however, is not sufficient since it must be reconciled with other physical properties of the medium. To understand the chemical composition of these environments and to direct future molecular searches in the framework of future astronomical observatories, it is important to untangle the detailed chemical reactions and processes leading to the formation of new molecules in extraterrestrial environments. These data, together with quantum chemical calculations, will establish credible chemical models of interstellar, planetary, and stellar environments that are imperative to predict the existence of distinct molecules in extraterrestrial environments, thus guiding future astronomical searches of hitherto unobserved molecular species.

### 4.3. Solid Materials

In order to properly decipher the mechanisms that occur in interstellar cloud environments, laboratory studies of dust and ices are required (Tielens 2005). Studies of silicate and carbonaceous dust precursor molecules and grains are needed as are studies of the dynamical interaction between dust and its environment (including radiation and gas). Studies of dust and ice provide a clear connection between astronomy within and beyond the solar system. Astronomical observations and supporting laboratory experiments from the X-ray domain to the sub-millimeter regions are of paramount importance for studies of the molecular and dusty universe. Observations at infrared and sub-millimeter wavelengths penetrate the dusty regions and probe the processes occurring deep within them. These wavelengths provide detailed profiles of molecular transitions associated with dust. Because of its importance, astronomical observations centered on this wavelength region will chart star and planet formation in the Milky Way, the galactic life cycle of the elements, and the molecular and dusty universe.

Mid-IR spectra of individual objects such as H II regions, reflection nebulae, and planetary nebulae as well as the general interstellar medium are dominated by a set of emission features due to large aromatic molecules (a.k.a "PAH bands"). Studies of the spectral characteristics of such molecules and their dependence on molecular structure and charge state are of key importance for our understanding of this ubiquitous molecular component of the ISM. At long wavelengths, the continuum dust opacity is uncertain by an order of magnitude. IR spectral features of interstellar dust grains are used to determine their specific mineral composition, hence their opacities, which determine inferred grain temperatures and the masses of dusty objects, including the interstellar medium of entire galaxies. Emission bands from warm astronomical environments such as circumstellar regions, planetary nebulae, and star-forming clouds lead to the determination of the composition and physical conditions in regions where stars and planets form. The laboratory data essential for investigations of dust include measurements of the optical properties of candidate grain materials (including carbonaceous and silicate materials, as well as metallic carbides, sulfides, and oxides) as a function of temperature. For abundant materials (e.g., forms of carbon such as PAHs), the measurements should range from gas-phase molecules to nanoparticles to bulk materials. The IR spectral region is critical for the identification of grain composition, but results are also required for shorter wavelengths (i.e., UV), which heat the grains. The UV spectral region contains features of important large interstellar molecules, such as organic species which carry the IR emission bands and diffuse interstellar bands (DIBs) and that may be related to the origin of life. Studies of the UV characteristics of such molecules and their dependence on molecular structure and charge state are of importance for our understanding of this molecular component of the ISM. The lack of experimental data in this spectral region has hampered progress in theoretical studies as well as the interpretation of astronomical data. UV spectra are uniquely capable of identifying specific molecules, in contrast with the less specific transitions observed in the IR. Laboratory studies provide spectroscopy of large organic molecules and their ions. This work must be complemented by quantum theory calculations so that the laboratory data are properly interpreted.

### 4.4. Plasma Physics

Much of the Galactic neighborhood is in the plasma state. Modeling of plasma systems is beset with difficulties - none of turbulence, radiative energy transfer, collisionless shocks, and complex magnetic geometries can now be accurately modeled over scales of interest. As a result, laboratory plasma physics has much to contribute to our understanding of the corresponding dynamics.

Plasma hydrodynamic experiments can explore well-scaled dynamics of systems of interest, building on recent experience in which an experiment (Klein et al. 2003) has been used (Hwang et al. 2005) to directly interpret Chandra results observing a clump being destroyed by an SNR blast wave. This will contribute to the interpretation of data on the interactions of blast waves, shocks, and winds with interstellar structure from future observatories. Laboratory environments can produce radiating systems having dimensionless parameters in regimes relevant for example to shock breakout from supernovae, which will be relevant to the anticipated great increase in such data and to the simulation codes used to interpret it. To benchmark the models that are needed to interpret data from the photoionized regions mentioned above, one can irradiate a plasma volume containing relevant species with an intense x-ray source (e.g., Foord et al. 2006). In accretion disks, the magnetorotational instability (MRI) may provide the needed angular momentum transport. The MRI will be produced and studied in the laboratory during the next decade. This will prove to be of value in benchmarking the newly complex MHD codes needed to better understand phenomena such as stellar outflows, supernovae, and accretion disks. Other experiments that will contribute to such understanding include those on magnetic reconnection, which is already very active (e.g., Ren et al. 2005), on turbulent dynamos, which is poised to become active, and on collisionless shocks, which is feasible but requires progress with experimental facilities.

### 4.5. Nuclear Physics

Major uncertainties remain in our understanding of the nuclear physics governing massive-star evolution. The most critical parameter in the hydrostatic evolution of supernova progenitors is the $^{12}C(\alpha,\gamma)$ cross section, which determines the relative masses of the C and O shells, and thus has a major influence on supernova nucleosynthesis. There are several experimental efforts underway to further constrain this cross section, which is complicated due to the contributions of sub-threshold resonances.

The path of the *r*-process is through heavy, neutron-rich nuclei unknown on Earth, but which form the equilibrium state of nuclear matter in such astrophysical explosions. One of the main goals of the Facility for Rare Ion Beams (FRIB), a planned new Department of Energy accelerator, will be to produce these isotopes for the first time, so that their masses and beta decay lifetimes can be measured. The goal will be to characterize the nuclear physics well enough that the nucleosynthetic output of supernovae can then be used to constrain aspects of the explosion, such as the dynamic timescale for the ejecta and the freeze-out radius – just as Big Bang nucleosynthesis has been used to constrain conditions in the early Universe. These studies will also provide clues to the nature of the first stars in the Galaxy. As for the *s*-process, there are uncertainties associated with long-lived, but unstable nuclei. Current plans involve harvesting isotopes with FRIB and providing these targets for use at other facilities. Additional progress is anticipated with the Spallation Neutron Source at Oak Ridge National Laboratory.

### 5. Four central questions and one area with unusual discovery potential

### 5.1 Four central questions
- How have the abundances of chemical elements evolved over cosmic time?
- How are the Galactic disk and halo coupled through a dynamic interstellar medium?
- What is the relationship between the Galactic Center and active galactic nuclei throughout cosmic time?
- What is the life cycle of cosmic carbon and what are the chemical steps leading to life on Earth?

### 5.2 One area with unusual discovery potential
- The field of astrobiology holds the promise of discovering how life arose on Earth and how likely it is elsewhere in our Galactic neighborhood.

## 6. Postlude

Laboratory astrophysics and complementary theoretical calculations are part of the foundation for our understanding of GAN and will remain so for many generations to come. From the level of scientific conception to that of the scientific return, it is our understanding of the underlying processes that allows us to address fundamental questions in these areas. In this regard, laboratory astrophysics is much like detector and instrument development; these efforts are constantly necessary to maximize the scientific return from astronomical observations.

## References

Bregman, J.N. 1980, ApJ, 236, 577.
de Avillez, M.A., & Breitschwerdt, D. 2004, A&A, 425, 899.
Den Hartog, D.A. et al. 2006, ApJS, 167, 292.
Ganguly, R. et al. 2005, ApJS, 157, 251.
Foord, M. E. et al. 2006, JQSRT, 99, 712.
Gaensler, B. M., and Slane, P. O. 2006, Ann. Rev. A&A, 44, 17.
Heil, M. et al. 2008, Phys. Rev. C, 78, 025802.
Hwang, U. et al. 2004, ApJ, 615, L117.
Hwang, U. et al. 2005, ApJ, 635, 355.
Klein, R. I. et al. 2003, ApJ, 583, 245.
Lawler, J.E. et al. 2008, ApJS, 178, 71.
Lodders, K. 2003, ApJ, 591, 1220.
Porquet, D. et al. 2008, A&A, 488, 549.
Ren, Y., et al. 2005, Phys. Rev. Lett., 95, 055003.
Salama, F. 1999 in Solid Interstellar Matter: The ISO Revolution, d'Hendecourt et al. eds.,
    EDP Sciences, Springer-Verlag.
Scott, P. et al. 2009, ApJ, 691, L119.
Shapiro, P.R., & Field, G.B. 1976, ApJ, 205, 762.
The, L-S et al. 2007, ApJ, 655, 1058.
Tielens, A. G. G. M. 2005, The Physics and Chemistry of the Interstellar Medium,
    Cambridge University Press.
Tüllman, R. et al. 2006, A&A, 448, 43.
Williams, R.J. et al. 2007, ApJ, 665, 247.
Yusef-Zadeh, F. et al. 2008, ApJ, 683, L147.

**Appendix III: New Discoveries in Galaxies across Cosmic Time through Advances in Laboratory Astrophysics**

### 1. Introduction

As the Galaxies across Cosmic Time (GCT) panel is fully aware, the next decade will see major advances in our understanding of these areas of research. To quote from their charge, these advances will occur in studies of "the formation, evolution, and global properties of galaxies and galaxy clusters, as well as active galactic nuclei and QSOs, mergers, star formation rate, gas accretion, and supermassive black holes."

Central to the progress in these areas are the corresponding advances in laboratory astrophysics that are required for fully realizing the GCT scientific opportunities within the decade 2010-2020. Laboratory astrophysics comprises both theoretical and experimental studies of the underlying physics that produce the observed astrophysical processes. The 5 areas of laboratory astrophysics that we have identified as relevant to the CFP panel are atomic, molecular, solid matter, plasma, nuclear, and particle physics.

In this white paper, we describe in Section 2 some of the new scientific opportunities and compelling scientific themes that will be enabled by advances in laboratory astrophysics. In Section 3, we provide the scientific context for these opportunities. Section 4 briefly discusses some of the experimental and theoretical advances in laboratory astrophysics required to realize the GCT scientific opportunities of the next decade. As requested in the Call for White Papers, Section 5 presents four central questions and one area with unusual discovery potential. Lastly, we give a short postlude in Section 6.

### 2. New scientific opportunities and compelling scientific themes

The X-ray spectra of many active galactic nuclei (AGNs) show numerous absorption features produced by ionized gas in an outflowing wind. The significance is still being debated of these winds for the AGN central engine or for the host galaxy in terms of mass, energy, and momentum. These outflows may also play a central role in cosmological feedback and in the metal enrichment of the intergalactic medium (Holczer et al. 2007).

Cooling flows in the cores of clusters of galaxies have been a subject of intense study for several decades and the details of the cooling process remain highly uncertain. *XMM-Newton* spectroscopic observations "exhibit strong emission from cool plasma at just below the ambient temperature $T_0$, down to $T_0/2$, but also exhibit a severe deficit of emission relative to the predictions of isobaric cooling-flow model at lower temperatures ($<T_0/3$). In addition, the best-resolved spectra show emission throughout the entire X-ray temperature range, but increasingly less emission at lower temperatures than the cooling-flow model would predict. These results are difficult to reconcile with simple prescriptions for distorting the emission measure distribution, e.g., by including additional heating or rapid cooling flows" (Peterson et al. 2003).

*XMM-Newton* observations of elliptical galaxies have "derived metal abundances which are higher than have usually been inferred from prior, lower spectral resolution observations [of NGC 4636] but are still incompatible with conventional chemical-enrichment models of elliptical galaxies. In addition, [these observations] are incompatible with standard cooling-flow models for this system" (Xu et al. 2003).

One of the big remaining needs for X-ray astronomy is spatially-resolved observations at high spectroscopic resolution for galaxies, clusters of galaxies, and supernova remnants (SNRs). Such observations promise to provide a wealth of new information about the properties of these objects.

Recent observations of galaxies in the distant Universe reveal the presence of interstellar material much like that seen in the Milky Way. For instance, HCN emission, a dense gas tracer, is seen in sub-millimeter galaxies and QSOs (Gao et al. 2007). Even the more widespread emission associated with PAHs is observed in distant galaxies (e.g., Rigby et al. 2008). As more detailed studies become available over the next decade, the correspondence between the emission from local gas clouds and those seen in galaxies from earlier epochs will be more clearly discerned.

## 3. Scientific context

The upcoming decade promises numerous opportunities for progress in the areas of GCT. There are a multitude of current and planned ground-based and satellite observatories including (but not limited to) the Arizona Radio Observatory (ARO), the Atacama Large Millimeter/submillimeter array *(ALMA)*, *Chandra*, the Gemini telescopes, *Herschel*, the Hobby-Eberly telescope *(HET)*, *Hubble*, the Institut de RadioAstronomie Millimétrique *(IRAM)*, the James Clerk Maxwell Telescope *(JCMT)*, *JWST*, *Keck*, *Kepler*, the Large Binocular Telescope *(LBT)*, Owens Valley Radio Observatory *(OVRO)*, *Magellan*, MMT Observatory *(MMTO)*, *NEXT*, *SOFIA*, *Spitzer*, the Southern African Large Telescope *(SALT)*, the Submillimeter Array *(SMA)*, *Subaru*, the Swedish-European Space Organization Submillimeter Telescope *(SEST)*, the *Very Large Array (VLA)*, and *XMM-Newton*. Studies from the radio to X-rays contribute to our overall understanding of the physical and chemical processes important in GCT. Interpreting these data will require an accurate understanding of the underlying microphysics that produces the observed spectra.

Recent *Chandra* and *XMM-Newton* spectroscopic observations of AGNs have provided a wealth of data and discovery in these objects. From the presence of X-ray absorbing outflows to the existence of new phases of gas in these outflows as detected through absorption by Fe M-shell unresolved transition array (UTA) features, astrophysicists have deepened their understanding of these objects while simultaneously raising new questions. Answering these questions is one of the major drives behind the planned launch of the *International X-Ray Observatory (IXO)*.

X-ray spectroscopic observations for cooling-flows and metal abundances in clusters of galaxies and for elliptical galaxies are a major impetus behind future X-ray satellite observatories. Particularly important are spatially and spectrally resolved observations such as will be provided by the planned microcalorimeter on the Japanese satellite mission *NEXT*. Such observations will allow researchers to use spectroscopic diagnostics to develop a spatially resolved map for the physical properties of the cooling-flows and the metal abundances. The expected microcalorimeter observations of galaxies and SNRs also are expected to revolutionize our understanding of these sources.

In our Galaxy, dense molecular clouds are the sites of star formation. They appear to be involved with star formation in the Galactic Center (Doeleman et al. 2008), even in the presence of phenomena associated with a supermassive black hole. Moreover, PAH emission is ubiquitous in clouds within the disk of our Galaxy. Now that molecular emission from dense clouds and PAH emission are seen is distant galaxies, it is important to examine whether the inferred properties derived locally apply to galaxies at a much earlier cosmic time. The new facilities nearing completion will allow more detailed comparisons than ever before.

## 4. Required advances in Laboratory Astrophysics

Advances particularly in the areas of atomic, molecular, solid matter, plasma, nuclear, and particle physics will be required for the scientific opportunities described above. Here we briefly discuss some of the relevant research in each of these 6 areas of laboratory astrophysics. Experimental and theoretical advances are required in all these areas for fully realizing the GCT scientific opportunities of the next decade.

### 4.1. Atomic Physics

Reliably characterizing the speed of the AGN outflows requires accurate experimental and theoretical wavelengths for the X-ray absorption lines used to measure the outflow velocities.

The shapes, central wavelengths, and equivalent widths of the AGN UTAs can be used to diagnose the properties of the AGN warm absorbers (Behar et al. 2001). However models that match K-shell absorption features from 2nd and 3rd-row elements cannot reproduce correctly the observed UTAs from the fourth-row element iron (Netzer et al. 2003). The problem is due in part to the lack of reliable low temperature dielectronic recombination data for Fe M-shell ions (Netzer 2004, Kraemer et al. 2004).

Spectroscopic diagnostics using lines from He-like ions and line ratios from other isoelectronic sequences can give temperature measurements independent of ionization state. Because these results do not require a model for the charge state distribution of the gas, they are robust for both ionization equilibrium and more importantly for cases of non-equilibrium ionization. For clusters of galaxies, it is known that resonance scattering of hydrogenic-like Fe Lyman α and other resonance series lines is important. Hence, given the inferred temperature, the Lyman α and other resonance series line ratios can be used to derive an independent measure of the column density for clusters. Accurate theoretical line ratios are crucial for these temperature measurements. Uncertainties on the order of 20% can result in a factor of 2 error in the inferred temperature (Chen et al. 2006).

Reliably determining the electron and ion temperatures, densities, emission measure distributions, and ion and elemental abundances of cosmic atomic plasmas requires accurate fractional abundance calculations for the different ionization stages of the various elements in the plasma (i.e., the ionization balance of the gas). Since many of the observed sources are not in local thermodynamic equilibrium, in order to determine the ionization balance of the plasma one needs to know the rate coefficients for all of the relevant recombination and ionization processes. Radiative recombination, dielectronic recombination, and charge transfer are the primary recombination processes to consider. Ionization includes that due to photons, electrons, and from charge transfer. Considering all the ions and levels that need to be taken into account, it is clear that vast quantities of data are needed. Generating them to the accuracy required for astrophysics pushes theoretical and experimental methods to the edge of what is currently achievable and often beyond. For this reason progress is slow and every 5-10 years a group of researchers takes it upon themselves to survey the field, evaluate and compile the relevant rate coefficients, and provide the most reliable ionization balance calculations possible at that time (Savin 2005, Bryans et al. 2006). Such papers are among the most highly cited laboratory astrophysics publications (e.g., Mazzotta et al. 1998 with over 500 citations).

Analyzing and modeling spectra begin with identifying the observed lines that may be seen in emission or absorption. This requires accurate and complete wavelengths across the electromagnetic spectrum. The next step toward understanding the properties of an observed cosmic source depends on accurate knowledge of the underlying atomic processes producing the observed lines. Oscillator strengths and transition probabilities are critical to a wide variety of temperature and abundance studies from infrared to X-ray wavelengths. Many existing data for the heavier elements are still notoriously unreliable.

Line ratios, a key diagnostic, involve knowledge of collision strengths and related phenomena. Rate coefficients for electron impact excitation approaching 10% accuracy are necessary for the most important line ratio diagnostics yielding temperature, optical depth, density, and abundance. Proton impact excitation is important because ions in hot post-shock material decouple from radiatively cooling electrons and may remain hot enough to produce line emission through collisional impact. Furthermore, electron impact ionization (EII) data are highly suspect. Recommended data derived from the same scant set of measurements and calculations can differ by factors of 2 to 3. Much of the published experimental data include contributions from an unknown metastable fraction in the ion beams used. Little data also exist for three-body recombination, the time reverse of EII, which is important in high density plasmas.

### 4.2. Molecular Physics

High-resolution laboratory spectroscopy is absolutely essential in establishing the identity and abundances of molecules observed in astronomical data. Given the advancements in detector technologies, laboratory measurements need to have a resolving power higher than the astronomical instruments. This is essential to interpret ubiquitous interstellar spectral features such as the IR emission bands ("PAH bands") as well as the dense molecular spectra found in cloud cores, both of which hold keys to our understanding of the molecular universe (Salama 1999). It is also crucial in making the link between interstellar molecules in the gas phase and dust grains in the solid phase, a key phase in our understanding of the evolution of interstellar clouds. For molecular data obtained from astronomical observations to be of practical use, accurate assignments of observed spectral features are essential. The problem here is two-fold. First, the transitions of known molecules need to be assigned in these spectra, including higher energy levels and new isotopic species. Second, the spectra of undiscovered species that promise to serve as important new probes of astronomical sources need to be identified.

The spectroscopic study of interstellar molecules, many of which are complex structures that cannot be produced in large abundance in the laboratory, requires the development and application of state-of-the-art ultra-sensitive spectroscopic instruments. Detecting the possible presence of a species, however, is not sufficient since it must be reconciled with other physical properties of the medium. To understand the chemical composition of these environments and to direct future molecular searches in the framework of future astronomical observatories, it is important to untangle the detailed chemical reactions and processes leading to the formation of molecules in extraterrestrial environments. These data, together with quantum chemical calculations, will establish credible chemical models.

## 4.3. Solid Material

In order to properly decipher the mechanisms that occur in interstellar cloud environments, laboratory studies of dust and ices are required (Tielens 2005). Studies of silicate and carbonaceous dust precursor molecules and grains are needed as are studies of the dynamical interaction between dust and its environment (including radiation and gas). Astronomical observations and supporting laboratory experiments from the X-ray domain to the sub-millimeter regions are of paramount importance for studies of the molecular and dusty universe. Observations at infrared and sub-millimeter wavelengths penetrate the dusty regions and probe the processes occurring deep within them. These wavelengths provide detailed profiles of molecular transitions associated with dust.

Mid-IR spectra of individual objects such as H II regions, reflection nebulae, and planetary nebulae as well as the general interstellar medium are dominated by a set of emission features due to large aromatic molecules (a.k.a "PAH bands"). Studies of the spectral characteristics of such molecules and their dependence on molecular structure and charge state are of key importance for our understanding of this ubiquitous molecular component of the ISM. At long wavelengths, the continuum dust opacity is uncertain by an order of magnitude. IR spectral features of interstellar dust grains are used to determine their specific mineral composition, hence their opacities, which determine inferred grain temperatures and the masses of dusty objects, including the interstellar medium of entire galaxies. Emission bands from warm astronomical environments such as star-forming clouds lead to the determination of the composition and physical conditions in these regions. The laboratory data essential for investigations of dust include measurements of the optical properties of candidate grain materials (including carbonaceous and silicate materials, as well as metallic carbides, sulfides, and oxides) as a function of temperature. For abundant materials (e.g., forms of carbon such as PAHs), the measurements should range from gas-phase molecules to nanoparticles to bulk materials. The IR spectral region is critical for the identification of grain composition, but results are also required for shorter wavelengths (i.e., UV), which heat the grains. The lack of experimental data in this spectral region has hampered progress in theoretical studies as well as the interpretation of astronomical data.

## 4.4. Plasma Physics

Plasma physics at high energy density can produce photoionized plasmas that can benchmark models of AGN and QSO absorbers. This is accomplished by producing an intense x-ray source that can irradiate a plasma volume containing relevant species, and measuring the ionization balance and other properties that result. Work in this direction has begun (Foord et al. 2006; Wang et al. 2008). Much more will be possible in the coming decade as higher-energy facilities come online. Another contribution is to produce environments with intense magnetic field such as exist in the vicinity of AGNs and QSOs. Understanding the behavior of charged particles in intense magnetic fields is important for star formation, gas accretion, feedback, and supermassive black holes. Such fields can also alter transition energies affecting the analysis of spectra from the vicinity of highly magnetized plasma. Gigagauss fields, large enough to have significant effects, are produced by current-generation intense lasers (Wagner et al. 2004); the next generation will reach 10 Ggauss.

## 4.5 Nuclear Physics

Experimental nuclear lab cross sections are helping us to determine heavy element abundances, particularly those from the r-process. Those abundance patterns in Galactic stars are "solar like", indicating surprisingly that the same types of nuclear processes have been occurring over Gyr in our Galaxy. Even more surprising observational studies of stars in distant galaxies, in for example Damped Lyman Alpha Systems, and in nearby galaxies also show the same solar system pattern for heavy elements. This points to a commonality in nucleosynthesis in the earliest stars in other galaxies (Cowan 2007).

A recent, much improved measurement at the underground LUNA facility, Gran Sasso, Italy, of the cross section for $^{14}N(p,\gamma)$ has altered the age estimates of globular clusters, increasing them by about 0.9 Gyr. Further improvements are anticipated. This reaction sets the clock for the CN cycle.

## 4.6 Particle physics

Galaxy formation rests on an initial set of seeds which represent a departure from homogeneity in the Universe. The amplitude of these perturbations is constrained by the observations of power spectrum in the cosmic microwave background (CMB) radiation. It is well known that these perturbations (thought to have been produced in the very early Universe during inflation) would not have sufficient time to grow in a baryon dominated universe. The accepted solution to this problem is dark matter. If dark matter is not coupled to radiation, then the perturbations in the dark matter can begin to grow under the influence of gravity before the epoch of the last scattering of the CMB. As a consequence, when atoms finally become neutral and are free to collapse into large scale structures, they already see well formed potentials in dark matter. In this way, galaxies and clusters of galaxies are seeded by existing structure in the dark matter. The study of dark matter in a multitude of laboratory experiments will help shed light on the important process of galaxy formation.

## 5. Four central questions and one area with unusual discovery potential

### 5.1 Four central questions

- What is the nature of the AGN warm absorber outflows?
- What is the correct physical description for cooling flows in galaxy clusters?
- What are the metal abundances and cooling-flow properties for elliptical galaxies?
- What is the correspondence between star formation and activity in our Galaxy and what is seen in the distant Universe?

### 5.2 One area with unusual discovery potential

- New facilities will provide the data needed to translate knowledge gained on our Galaxy and its neighbors with an understanding of the distant Universe.

## 6. Postlude

Laboratory astrophysics and complementary theoretical calculations are part of the foundation for our understanding of GCT and will remain so for many generations to come. From the level of scientific conception to that of the scientific return, it is our understanding of the underlying processes that allows us to address fundamental questions in these areas. In this regard, laboratory astrophysics is much like detector and instrument development; these efforts are necessary to maximize the scientific return from astronomical observations.

## References

Behar, E., et al. 2001, Astrophys. J., 563, 504.
Bryans, P., et al. 2006, Astrophys. J. Suppl. Ser., 167, 343.
Chen, G. X., et al. 2006, Phys. Rev. A, 74, 042709.
Cowan, J. 2007, Nature, 448, 29.
Doeleman, S. S., et al. 2008, Nature, 455, 78.
Foord, M. E., et al. 2006, J. Quant. Spectrosc. Radiat. Transfer, 99, 712.
Gao, Y., et al. 2007, Astrophys. J., 660, L93.
Holczer, T., et al. 2007, Astrophys. J., 663, 799.
Kraemer, S., et al., 2004, Astrophys. J., 604, 556.
Mazzotta, P., et al. 1998, Astron. Astrophys. 133, 403.
Netzer, H., et al., 2003, Astrophys. J., 599, 933.
Netzer, H. 2004, Astrophys. J., 604, 551.
Peterson, J. R., et al. 2003, Astrophys. J., 590, 224.
Rigby, J. R. et al. 2008, Astrophys. J., 675, 262.
Salama, F. 1999 in Solid Interstellar Matter: The ISO Revolution, d'Hendecourt et al. eds.,
Savin, D. W. 2005, AIP Conf. Proc., 774, 297.
Tielens, A. G. G. M. 2005, The Physics and Chemistry of the Interstellar Medium,
    Cambridge University Press.
Wagner, U., et al. 2004, Phys. Rev. E, 70, 026401.
Wang, F. L., et al. 2008, Phys. Plasmas, 15, 073108.
Xu, H., et al. 2003, Astrophys. J., 579, 600.

**Appendix IV: New Discoveries in Planetary Systems and Star Formation through Advances in Laboratory Astrophysics**

**1. Introduction**

As the panel on Planetary Systems and Star Formation (PSF) is fully aware, the next decade will see major advances in our understanding of these areas of research. To quote from their charge, these advances will occur in studies of "solar system bodies (other than the Sun) and extrasolar planets, debris disks, exobiology, the formation of individual stars, protostellar and protoplanetary disks, molecular clouds and the cold ISM, dust, and astrochemistry."

Central to the progress in these areas are the corresponding advances in laboratory astrophysics which are required for fully realizing the PSF scientific opportunities in the decade 2010-2020. Laboratory astrophysics comprises both theoretical and experimental studies of the underlying physics and chemistry which produce the observed spectra and describe the astrophysical processes. We discuss four areas of laboratory astrophysics relevant to the PSF panel: atomic, molecular, solid matter, and plasma physics.

Section 2 describes some of the new opportunities and compelling themes which will be enabled by advances in laboratory astrophysics. Section 3 provides the scientific context for these opportunities. Section 4 discusses some experimental and theoretical advances in laboratory astrophysics required to realize the PSF scientific opportunities of the next decade. As requested in the Call for White Papers, we present in Section 5 four central questions and one area with unusual discovery potential. We give a short postlude in Section 6.

**2. New scientific opportunities and compelling scientific themes**

In the upcoming decade, we will make progress in understanding the evolution from clouds of gas and dust to systems of stars and planets. Studies of protostellar and protoplanetary disks and debris disks will inspire new ideas to explain how planets form and migrate within young systems. We will not only continue to discover new and interesting planets but we will also begin to characterize and classify the properties of these planets and their atmospheres and cores. The discoveries of earth-like planets and other planets in habitable zones will challenge us to address the origin of life. We expect to solve long-standing mysteries, such as the nature of the diffuse interstellar bands (DIBs), believed to be the manifestation of an unidentified reservoir of organic material. Discovery of the carriers of these bands promises to revolutionize our understanding of organic chemistry in the interstellar medium (ISM).

Perhaps the most exciting new opportunities will come in the area of the origins of life. The cosmic pathway to life begins in interstellar gas clouds where atomic carbon is "fixed" into molecules, thereby initiating the synthesis of the complex organic species that are eventually sequestered on planets. These reactions initiate not only the formation of organic molecules in the cosmos, but also provide some of the first threads knitting atoms and molecules into solid material. Such processes are critical for the eventual formation of planets and may determine a major component of the organic chemistry that is present on their young surfaces. To place the new worlds we discover in proper context requires that we understand the chain of events that dictates their potential chemical inventories.

## 3. Scientific context

With a multitude of current and planned space-based and ground-based telescopes, including *Spitzer*, *Hubble*, *Chandra*, *XMM-Newton*, *Suzaku*, *Herschel*, *SOFIA*, K*epler*, Keck, Magellan, ARO, the MMT, Hobby-Eberly, SALT, LBT, the Gemini telescopes, and eventually ALMA and *JWST*, the upcoming decade promises numerous opportunities for progress in the areas of planet and star formation. Studies from the radio to X-rays contribute to our overall understanding of the physical and chemical processes.

Understanding the process of accretion is key to astrophysics from the formation of the lowest mass stars to the growth of super-massive black holes. Star formation studies over the past decade have begun to recognize the importance of the interaction between a young star and its environment. X-ray and ultraviolet (UV) radiation produced by a low mass star can have profound effects on the ionization and heating of gas and dust in the disk (Glassgold et al. 2004; Jonkheid et al. 2004). The star's magnetic fields can extend to the disk and help to channel the accreting material. Flares are ubiquitous in young low-mass stars (Feigelson et al. 2007) and thus the inner disk experiences highly time-variable phenomena. In addition to disk-driven winds, magnetic accretion can generate jets and winds.

The explosion of exoplanet discoveries over the past decade suggests that the next decade will revolutionize our understanding of the formation of planets, their atmospheres and cores, and the rich diversity of planetary systems. The transit method of exoplanet searching, at the core of *Kepler*'s goal to discover other Earths, requires accurate and finely-tuned atmosphere models. The first direct studies of light emitted by an exoplanet itself (Deming et al. 2005; Charbonneau et al. 2005) estimate temperatures from simple models fit to broad-band infrared (IR) observations; water has now been observed spectroscopically on an exoplanet (Grillmair et al. 2008). In the next decade observations will continue to drive the sophistication of planet atmosphere models (e.g. winds from close-in Jupiters; Murray-Clay et al. 2008). Gas giants dominate our planetary system by mass and may be ubiquitous in stellar systems. A favored model for their formation requires a metallic core; however, it is not clear whether the heavy core survives, an issue which can be addressed by experimental studies of high density plasmas.

The evolution of gas and mineral dust in protoplanetary disks traces the formation process of planets and other objects. Comparisons of exoplanet observations with studies of objects in our own solar system will help to determine the common evolutionary histories (e.g. Bergin et al. 2007). Discoveries of smaller planets and multi-planet systems will shed light on the timescales for formation, migration, destruction, and the relative roles of dust and gas. The growth of dust grains in the presence of stellar radiation depends on the gas-to-dust ratio and the dust temperature which, in turn, control the chemistry. Gas and dust are coupled at the high densities found in the mid-plane of the disk, suggesting that some molecules are frozen onto grains. Disk dynamics can alter the properties of gas and dust through mixing.

The discoveries of over 140 different chemical molecules in interstellar gas, with the vast majority organic molecules, and the realization that obscuring dust pervades vast regions of the ISM have revealed the complexity of interstellar chemistry. The first stars are thought to have formed from primordial clouds where $H_2$ and HD controlled their cooling and collapse (Bromm et al. 2002). Subsequent stars and planetary systems have formed out of the most complex molecular environments deep inside cold gas and dust clouds, often obscured by hundreds of visual magnitudes of extinction. All the main functional groups known to organic chemists have now been observed in interstellar molecules, suggesting that interstellar chemistry contains the organic complexity seen on Earth. The next decade may establish the gas phase and grain chemistries of interstellar clouds that can lead to the origin of life.

## 4. Required advances in Laboratory Astrophysics

Spectroscopic studies are key to identifying the constituents of various environments, diagnosing the physical and chemical conditions, and testing models of the physical processes and chemical reactions. A strong program in laboratory astrophysics allows us to make steady progress in compiling databases of the spectral properties of neutral atoms and ions, molecules, and minerals, and to react quickly to address surprises from new observations. While laboratory astrophysics has contributed to many successes in astrophysics in many areas of ongoing research in PSF fields, we know that the linelists are not complete enough and that the rate coefficients are not accurate enough to take full advantage of the wealth of new data expected over the next decade. Moreover, models of astrophysical plasma processes should be tested experimentally to ensure that the assumptions are valid.

### 4.1. Atomic Physics

Understanding the star-disk interaction in protostellar and young stellar systems requires accurate knowledge of the atomic physics and fundamental spectroscopy across many wavebands. X-ray emission from an accretion shock has been identified by X-ray spectral line ratios indicative of high electron density (Kastner et al. 2002). Unprecedented accuracy has now been demonstrated for atomic theory and experiment (Chen et al. 2006), but much more work is needed to generate a comprehensive set of reliable diagnostics for temperature and density. Stellar X-ray and UV radiation can excite specific lines in the disk, such Fe K$\alpha$ (Imanishi et al. 2001) and [Ne II] 12.8 μm (Pascucci et al. 2007). The processes involved in understanding such emission include photoionization, radiative and dielectronic recombination, collisional excitation, X-ray fluorescence, Auger ionization, and charge exchange. Accurate line lists, oscillator strengths, atomic transition probabilities, and collisional rate coefficients are needed to develop new diagnostics from the IR to the X-ray.

Planet searches require accurate stellar atmosphere models for cross correlation with the observed spectra (e.g. Konacki et al. 2004). Characterization of planetary atmospheres places new demands on atmosphere models for accurate and complete atomic line opacities. Ultimately the accuracy of the derived parameters depends on the accuracy of the template spectra; where lines are missing in the template, the observed spectrum cannot be used.

## 4.2 Molecular Physics

High-resolution laboratory spectroscopy is absolutely essential in establishing the identity and abundances of molecules observed in astronomical data. For molecules already identified in astronomical sources, transitions including higher energy levels and new isotopic species are needed as we move into new spectral regions and higher resolving powers. Furthermore, the spectra of as yet undiscovered species that promise to serve as important new probes of astronomical sources need to be identified. For example, the recent astronomical discovery of the negative ion $C_6H^-$ (McCarthy et al. 2006) was driven by laboratory work. We anticipate that additional discoveries of anions in the coming decade will challenge the existing view of interstellar chemistry.

New line diagnostics of disk heating, cooling, and ionization are expected from continuing laboratory work on abundant molecules. For example, submillimeter (sub-mm) lines of CO from high rotational levels have recently been observed in the outer disk of a young star, requiring additional heating such as by X-rays (Qi et al. 2006); numerous UV lines of $H_2$ were observed in the same system, pumped by X-ray or UV radiation (Herczeg et al. 2002).

The "unidentified line" problem needs to be addressed by systematic measurements of a variety of molecules. The DIBs were first observed in 1919, but despite many decades of intense efforts by laboratory spectroscopists and astronomers, the molecular carriers of these bands remain a mystery. Mid-IR spectra of individual objects such as H II regions, reflection nebulae, and planetary nebulae as well as the general ISM are dominated by a set of emission features due to large aromatic and aliphatic molecules (the so-called PAH bands). Studies of the IR spectroscopy of such molecules and their characteristic features which depend on molecular structure and charge state are of key importance for our understanding of this ubiquitous molecular component of the ISM. Millimeter and sub-mm spectra contain hundreds of unidentified gas-phase lines. Intense laboratory work continues to be a vital component in unlocking the mystery of the DIBs, the PAH emission bands, and other observed, but unidentified, spectral features (Salama 1999). Missing molecular lines in the coolest stellar atmosphere models and in the atmosphere models being developed for planets create an acute problem for interpreting and exploiting astronomical spectra. Discovery of biological molecules from planetary atmospheres may occur in the next decade (see Robinson et al. 2008).

A vast number of molecules which are difficult to detect by standard means may exist in the ISM and stellar systems. The identification of most interstellar/circumstellar species is done through observations of emission spectra in the radio and millimeter, with an observational bias in favor of molecules with large dipole moments. A big challenge for observation is to find ways to detect molecules with small or zero dipole moments. Laboratory spectroscopy in other wavebands would be useful.

Understanding line profiles will also provide new tools for PSF studies. For example, pressure broadening of the strong resonance absorption lines of Na and K, observed in brown dwarfs (Burrows et al. 2001) and predicted in extra-solar gas giants, occurs through collisions with neutrals, including molecular hydrogen (Seager & Sasselov 2000). Quantum collision calculations, benchmarked by laboratory measurements, can provide accurate line profiles which can serve as diagnostics of the temperature and pressure (e.g. Zhu et al. 2005).

Interpreting spectral observations requires the use of sophisticated astrochemical models, which include the ionization and dissociation mechanisms (by radiation and cosmic rays) and the time-dependent history of the clouds. To understand the chemical composition of these environments and to direct future molecular searches in the framework of new astronomical observatories, it is important to untangle the detailed chemical reactions and processes leading to the formation of new molecules in extraterrestrial environments. Breakthroughs in our understanding of the molecular universe are limited by uncertainties in the underlying chemical data in these models (e.g. Glover et al. 2006). Laboratory spectroscopy is crucial for understanding interstellar molecule formation in the gas phase and on grains.

Of particular importance are data for reactions of neutral atomic carbon atoms with molecular ions which are critical in initiating interstellar organic chemistry. Theory is limited to classical methods, as fully quantum mechanical reactions for systems with four or more atoms are beyond computational capabilities now and for the foreseeable future. Existing laboratory experiments have produced ambiguous results owing to the extraordinary challenge of generating and characterizing atomic carbon beams with no internal energy. The spectroscopic study of such molecules, many of which are complex structures that cannot be produced in large abundance in the laboratory, requires the development and application of state-of-the-art ultra-sensitive spectroscopic instruments. Tracking the cosmo-chemical pathway towards life requires cross-disciplinary work at the intersection of astronomy, chemistry, biology, and physics.

### 4.3. Solid Material

Spectroscopy provides unique information on stellar outflows and protoplanetary disks, accreting envelopes, the surrounding medium, the radiation fields, and planetary atmospheres. To be useful, however, this tool requires an ability first to identify the lines of gas phase molecules and solid state band features of grains, and then to explain their appearance using reliable physical excitation constants. Laboratory measurements supplemented by theoretical modeling provide this essential information.

An example of the problem of identification is the emission feature near 21 μm, originally discovered in four objects with large far IR excesses due to a circumstellar dust envelope surrounding a carbon-rich central star (Kwok et al. 1989). A similar feature, but with varying wavelength centroid, has been observed with Spitzer in the Carina nebula, the supernova remnant Cas A, and some H II regions and is likely associated with dust grains composed of FeO and a mix of silicates (e.g. Rho et al. 2008).

Laboratory measurements of molecules and grains at different temperatures and under various radiation environments are needed to inform the search for specific elements. For example, oxygen is the third most abundant element in the universe, and early models of gas clouds predicted that O I, $O_2$, and $H_2O$ should be relatively abundant. But observations over the past decade by SWAS, Odin, ISO, and other instruments found that $H_2O$ is about 100 times less abundant than expected in star forming regions; $O_2$ is similarly scarce. Oxygen plays a crucial role in the chemistry of star forming regions: pairing with carbon to make the crucial species CO for cooling the clouds (and thus facilitating cloud collapse), acting as a key ingredient in the chemical pathways that make most complex organic species, and by joining with silicon or hydrogen on the dust grains to form complex molecules. Recent calculations suggest that the missing oxygen may be in the form of water ice.

Studies of silicate and carbonaceous dust precursor molecules and grains are required as are studies of the interaction between dust and its environment, including radiation and gas (Tielens 2005). Studies of dust and ice provide a clear connection between astronomy within and beyond the solar system. Astronomical observations and supporting laboratory experiments over a wavelength range that extends from the X-ray domain to the UV, IR, and sub-mm are of paramount importance for studies of the molecular and dusty universe. Observations at IR and sub-mm wavelengths penetrate the dusty regions and probe the processes occurring deep within them. These wavelengths provide detailed profiles of molecular transitions associated with dust.

At longer wavelengths, the continuum dust opacity is uncertain by an order of magnitude. IR spectral features of interstellar dust grains are used to determine their specific mineral composition, hence their opacities, which determine inferred grain temperatures and the masses of dusty objects, including the ISM of entire galaxies. Emission bands from warm astronomical environments such as circumstellar regions, planetary nebulae, and star-forming clouds lead to the determination of the composition and physical conditions in regions where stars and planets form. The laboratory data essential for investigations of dust include measurements of the optical properties of candidate grain materials (including carbonaceous and silicate materials, as well as metallic carbides, sulfides, and oxides) as a function of temperature.

For abundant materials (e.g., forms of carbon such as PAHs), the measurements should range from gas-phase molecules to nanoparticles to bulk materials. The IR spectral region is critical for the identification of grain composition, but results are also required for the UV. Previous studies in the UV have focused on the only identified spectral feature (at 2200 Å), but COS on *Hubble* is expected to find UV spectral signatures for many materials, including specific individual aromatics.

### 4.4. Plasma Physics

Accretion from a disk orbiting a protostar requires efficient transport of angular momentum through the disk, but how this happens, and the extent to which it is dominated by nonlinear hydrodynamic turbulence or magnetohydrodynamic (MHD) turbulence via magnetorotational instability (MRI) (Balbus & Hawley 1998), has been much debated. Recent laboratory experiments show that pure hydrodynamic turbulence probably cannot explain the observed fast accretion (Ji et al. 2006). Laboratory jets are also providing tests of models, including studies on radiative cooling and collimation (Lebedev et al. 2002), instabilities (Ciardi et al. 2009), and dynamics and mixing (Foster et al. 2005).

How magnetic field changes its topology and releases its energy is an outstanding problem in plasma astrophysics (Biskamp 2000). Reconnection has been proposed as a mechanism for releasing the energy built up as a result of differential rotation between the star and the disk (van

Ballegooijen 1994). Only recently have reconnection models been tested experimentally. For example, the well-known Sweet-Parker model was tested quantitatively in the laboratory (Ji et al. 1998) 40 years after its birth, and the conjectured Hall effect has also now been successfully verified in the laboratory (Ren et al. 2005).

Knowledge of planetary interiors depends on models that incorporate the properties of the interior matter. As the interior pressure approaches and exceeds 1 Mbar (0.1 TPa, $10^{12}$ dynes/cm$^2$), the interior matter ionizes and becomes a high-energy-density plasma. There are important transitions involving molecular dissociation, ionization, Fermi degeneracy, and other effects under conditions where standard theoretical treatments for cold, condensed matter, or hot, ionized plasma are invalid. This is precisely the region of interest for planetary formation, structure and interior dynamics. Understanding the compressibility of the material is essential to develop structural models consistent with data. Understanding conductivity is essential to understand planetary magnetospheres. The existence of a metallic state of hydrogen has been confirmed in the laboratory (Weir et al. 1996), but the full insulator-to-conductor transition has yet to be explored. Understanding phase transitions and material mixing is essential to know how gravitational potential energy contributes to the energy balance. The anomalous radiation from nearby gas giant planets may be due to the separation, condensation, and raining downward of helium that is initially mixed with hydrogen (Guillot 1999). Alternatively, perhaps so-called pycnonuclear reactions (Ichimaru & Kitamura 1999) may release energy in dense, metallic hydrogen and deuterium.

In the next decade, further results are expected from the laboratory in the areas of MHD, jets and magnetic reconnection. Experimental facilities that can create and study the states of matter present in planetary interiors exist or are under construction both in the US and worldwide. What is required is to exploit and to extend the diagnostic capabilities that can quantify the detailed properties of matter in planetary interiors, including not only density and pressure but also conductivity, viscosity, and other characteristics.

### 5. Four central questions and one area with unusual discovery potential

### 5.1 Four central questions

- How does a young star interact with its environment?
- What are the characteristics of planetary systems and their atmospheres and interiors?
- How do the gas and dust in disks around stars evolve and how are they affected by disk dynamics, stellar radiation, and grain growth?
- What are the abundances, distribution and formation mechanisms of the chemical compounds found in the ISM?

### 5.2 One area with unusual discovery potential

- What is the cosmo-chemical pathway from atoms in interstellar space to life?

## 6. Postlude

Laboratory astrophysics and complementary theoretical calculations are part of the foundation for our understanding of PSF and will remain so for many decades to come. From the level of scientific conception to that of scientific return, our understanding of the underlying processes allows us to address fundamental questions in these areas. Laboratory astrophysics is constantly necessary to maximize the scientific return from astronomical observations.

## References

Balbus, S., & Hawley, J. 1998, Rev. Mod. Phys., 70, 1
Bergin, E. et al. 2007, in Protostars and Planets V, 751
Biskamp, D. 2000, Magnetic Reconnection in Plasmas, Cambridge U Press
Bromm, V. et al. 2002, ApJ, 564, 23
Burrows, A. et al. 2001, Rev. Mod. Phys., 73, 719
Charbonneau, D. et al. 2005, ApJ, 626, 523
Chen, G.-X. et al. 2006, Phys. Rev. A, 74, 42749
Ciardi, A. et al., ApJ, 691, L147
Deming, D. et al. 2005, Nature, 434, 740
Feigelson, E. et al. 2007, in Protostars and Planets V, 313
Foster, J. M. et al. 2005, ApJ, 634, L77
Glassgold, A. E. et al. 2004, ApJ, 615, 972
Glover, S. C. et al. 2006, ApJ, 640, 553
Grillmair, C. J. et al. 2008, Nature, 456, 767
Guillot, T. 1999, Science, 286, 7277
Herczeg et al. 2002, ApJ, 572, 310
Imanishi, K. et al. 2001, ApJ, 557, 747
Ichimaru, S. & Kitamura, H. 1999, Phys Plasma, 6, 2649
Ji, H. et al. 1998, Phys. Rev. Lett., 80, 3256
Ji, H. et al. 2006, Nature, 444, 343
Jonkheid, B. et al. 2004, A&A, 428, 511
Kastner, J. H. et al. 2002, ApJ, 567, 434
Konacki, M. et al. 2004, ApJ, 609, L37
Kwok, S. et al. 1989, ApJ, 345, L51
Lebedev, S. V. et al. 2002, ApJ, 564, 113
McCarthy, M. C. et al. 2006, ApJ, 652, L141
Murray-Clay, R. et al. 2008, ApJ, accepted
Pascucci, L. et al. 2007, ApJ, 663, 383
Qi, C. et al. 2006, ApJ, 636, 157
Ren, Y. et al. 2005, Phys. Rev. Lett., 95, 055003
Rho, J. et al. 2008, ApJ, 673, 271
Robinson, T. et al. 2008, AAS, DPS, 40, 1.04
Salama, F. 1999 in Solid Interstellar Matter: The ISO Revolution, Springer-Verlag
Seager, S. & Sasselov, D. 2000, ApJ, 537, 916
Tielens, A. 2005, The Physics and Chemistry of the Interstellar Medium, Cambridge U. Press
van Ballegooijen, A. 1994, Space Sci. Rev., 68, 299
Weir, S. T. et al. 1996, Phys. Rev. Lett., 76, 18601863
Zhu, C. et al. 2005, J. Phys. A, 71, 052710

**Appendix V: New Discoveries in Stars and Stellar Evolution through Advances in Laboratory Astrophysics**

## 1. Introduction

As the Stars and Stellar Evolution (SSE) panel is fully aware, the next decade will see major advances in our understanding of these areas of research. To quote from their charge, these advances will occur in studies of "the Sun as a star, stellar astrophysics, the structure and evolution of single and multiple stars, compact objects, SNe, gamma-ray bursts, solar neutrinos, and extreme physics on stellar scales."

Central to the progress in these areas are the corresponding advances laboratory astrophysics, required to fully realize the SSE scientific opportunities within the decade 2010-2020. Laboratory astrophysics comprises both theoretical and experimental studies of the underlying physics that produces the observed astrophysical processes. The 6 areas of laboratory astrophysics which we have identified as relevant to the CFP panel are atomic, molecular, solid matter, plasma, nuclear physics, and particle physics.

In this white paper, we describe in Section 2 the scientific context and some of the new scientific opportunities and compelling scientific themes which will be enabled by advances in laboratory astrophysics. In Section 3, we discuss some of the experimental and theoretical advances in laboratory astrophysics required to realize the SSE scientific opportunities of the next decade. As requested in the Call for White Papers, Section 4 presents four central questions and one area with unusual discovery potential. Lastly, we give a short postlude in Section 5.

## 2. Scientific context, opportunities, and compelling themes

Stellar structure has been of interest since at least the early 20$^{th}$ century work of Eddington and his contemporaries. It took a bit longer to begin to understand stellar evolution and the role of stars in the production of the elements. This, however, was only the start. With the advent of observatories that can detect energetic radiation from x-rays to gamma rays to neutrinos, millimeter radiation revealing molecular lines, our awareness of the scope of phenomena in stars, PNe and SNe has broadened vastly. The discovery of gamma ray bursts and other dramatic and unexpected events has stimulated great continuing interest. Meanwhile, simulations have made it possible to integrate our apparent understanding of stellar dynamics and have turned out to show how little we understand the details of stars and their behavior. Laboratory astrophysics can contribute to this knowledge in myriad ways, ranging from the measurement of specific cross sections, reaction rates, and spectral features needed to interpret data and model stars, to the examination of physical processes that cannot be approximated well in the simulations.

The quest to understand in quantitative detail the origin of the elements and Galactic chemical evolution remains an enduring theme. Multiple uncertainties exist regarding both the long-term structure and evolution of stars and the brief but essential explosion phase. We know far more about the interior of the sun than of any other star, thanks to helioseismology. One of the things we know at present is that the predicted location of the boundary of the solar convective zone disagrees with that observed by 13 standard deviations (Basu and Antia, 2008). We know that neutrinos provide the energy that powers core-collapse SNe, but how they do so remains mysterious. We know that SNe are asymmetric and that they develop complex internal structure, but not in detail how they do so. We know that stars typically are magnetized, but do not understand in detail the role of magnetic fields in their evolution.

A second theme is the role of stars, PNe and SNe as immediate drivers of nearby structure. As the star evolves, so does the stellar wind, introducing structured material around the star, for example producing the varied morphologies of PNe. The SNe is clearly not uniform and spherical, and its nonuniformities carry structure into the surrounding medium. In the longer term, both the shock and the structured material from the SNe interact with the surrounding medium. Then the structure we observe has contributions from both the star and the ISM.

A third theme is the exploration of the extreme physics present in SSE. The environment around various compact objects has magnetic fields strong enough to alter atomic structure, electron-positron pair plasmas, relativistic shock waves and jets, and photoionized plasmas. The emissions from such systems are often strongly altered from those that would be produced in less extreme environments. Interpreting them requires models of complex, extreme physics. Knowing the models are correct requires experiments in relevant regimes.

A fourth theme is the structure of stellar atmospheres above the photosphere. That stellar coronae are ~ 1000 times hotter than the underlying photospheres has long been known but not understood. A major clue is the observation in the sun that the abundances of coronal elements with a low first ionization potential are enhanced over their photospheric values. Such abundance fractionization is observed in many stars, but some stars exhibit no such effect or the reverse effect. Detailed models attribute these effects to ponderomotive interactions with Alfven waves (Laming 2009)

## 3. Required advances in Laboratory Astrophysics

### 3.1. Atomic Physics

*Line identification* from emission or absorption spectra is the first step in analyzing and modeling their origins. This requires accurate and complete wavelengths across the electromagnetic spectrum. Understanding the properties of an observed cosmic source also depends on accurate knowledge of the underlying atomic processes. Oscillator strengths and transition probabilities are critical to a wide variety of temperature and abundance studies. Quantum calculations of collisional excitation rate coefficients, accurate to better than 10% and benchmarked with experiments, are needed for diagnostics of coronal and shocked plasmas. Accurate flourescence yields for K alpha lines of iron group elements, such as Cr and Mn, may help to determine the detonation mechanism in SN explosions (Badenes et al. 2008). Many existing data for the heavier elements are still notoriously unreliable. The current limitations on the atomic data available for mid-Z elements make it difficult to determine the nature of the *r-*process.

*Modeling ionization structure* requires accurate data on many processes. Models of electron ionized gas need high-temperature, density-dependent dielectronic recombination (DR) rates, becoming possible with 3$^{rd}$ and 4$^{th}$ generation light sources. Electron-impact ionization data need to be improved; existing measurements and calculations are often scant. Charge exchange (CX) interactions with H, $H^+$, He and $He^+$ are important for many systems, but few modern calculations or laboratory measurements exist at the relevant temperatures. Data are also needed for elements such as Se and Kr in order to study nucleosynthesis in the progenitors of PNe.

Understanding photoionization needs contributions from both atomic physics and plasma physics. Models need low-temperature DR rates, for which theory is challenging and laboratory measurements with light sources may provide the only reliable data. One can also produce photoionized plasmas to benchmark models by using an intense x-ray source to irradiate a volume containing relevant species, and measuring the ionization balance and other properties (Foord et al., 2006). The coming decade will see much more such work.

*Stellar opacities* contribute significantly to the questions and discrepancies that remain in understanding stellar evolution, and are an area where atomic physics and plasma physics are both essential. "Pulsed power" devices have used axial currents to create plasmas at the densities and temperatures which are present in the solar convection zone. This has enabled direct measurement of the x-ray opacity of iron, showing some discrepancies from the values used in present-day solar models (Bailey et al., 2009). In the next decade experiments will produce and exploit plasmas corresponding to increasing depth in the sun and will explore a more complete range of elements and conditions.

*Integrated modeling* of stars and supernovae requires all the above. This has recently (Bryans et al. 2009) led to the most stringent spectroscopic constraints to date for models of abundance fractionization and coronal heating. Line emission of the various ions in SNe can be used to test explosion models (Laming and Hwang 2004) and nucleosynthesis models. The observed distribution of elements within a SNR can constrain the processes occurring during the explosion (Akiyama et al. 2002). This requires accurately taking into account line emission, non-equilibrium ionization, radiative cooling and hydrodynamics. The resulting constraints are only as good as the underlying atomic data in the models. Improved atomic data are needed in order to improve them.

### 3.2. Molecular Physics

*Identifying molecules and their abundances* absolutely requires high-resolution laboratory spectroscopy. Cool stellar atmospheres exhibit a range of molecular optical and infrared spectra. Circumstellar shells of evolved stars, both AGB and supergiants, are known to be rich in molecular compounds, as evidenced by millimeter and infrared observations. The carbon-rich envelope of the AGB star IRC+10216 is one of the richest molecular sources in the Galaxy. Even in PNe, molecular material persists. Understanding the underlying chemistry and its relationship to dust formation in the stellar envelope is crucial in evaluating mass loss from evolved stars. Studies of elemental and isotopic ratios in circumstellar molecules has also enabled observational tests of models of nucleosynthesis.

The spectroscopic study of circumstellar molecules, many which exhibit exotic structures that cannot be produced in large abundance in the laboratory, requires the development and application of state-of-the-art ultra-sensitive spectroscopic instruments. Construction and implementation of these instruments is costly and time-consuming, and data production from these devices cannot be turned on and off at will. Efficient utilization requires continued support. Detecting the presence of a species, however, is not sufficient since it must be reconciled with other physical properties of the surrounding medium. It is important to untangle the detailed chemical reactions and processes leading to the formation of new molecules in the complex environments of stellar atmospheres and ejecta. This requires the use of state-of-the-art experimental facilities. These data, together with quantum chemical calculations, will establish credible chemical models of stellar environments, thus guiding future astronomical searches of hitherto unobserved molecular species.

### 3.3. Solid Material

*Dust identification* and the understanding of dust evolution in stellar environments depend on having a systematic catalog of observable features useful for identification of specific dust components and molecular precursors. Laboratory spectroscopy of minerals and carbonaceous materials provide this information; multi-waveband studies are particularly useful for confirming identifications made from only a single feature. As an example, a feature near 21 microns observed with Spitzer in the Carina nebulaa, the SN remnant Cas A, and in some H II regions is likely associated with dust grains composed of FeO and a mix of silicates (e.g. Rho et al. 2008). The wavelength centroid varies amongst the objects observed, suggesting either coagulation or a difference in the mixture of components.

### 3.4. Plasma Physics

*Explosion hydrodynamics* is uncertain for several reasons, one of which is that calculations which fully resolve the turbulent dynamics will not be possible for several more decades. One can observe the dynamics in experiments that are well scaled to explosion conditions (Kuranz et al., 2009). Some experiments like the observations have found unanticipated outward penetration of inner material (Drake et al. 2004). The next decade will see further progress on this issue.

*Shock breakout* observations from SNe are now possible and will soon become common (e.g., Soderberg et al., 2008). The radiative shock that emerges is in a novel regime, not readily modeled even in dedicated computer codes. Experiments have produced radiative shocks in this regime, and will proceed during the next decade to benchmark computer simulations of such systems and to seek unanticipated dynamics.

*Extreme physics* is present in GRBs, which produce Lorentz factors of a few hundred and radiate strongly. Present-day relativistic lasers can produce the same Lorentz factors in systems of the same dimensionless scale (number of skin depths) as GRBs and only a very few orders of magnitude higher in density. This will enable experiments that are directly relevant to GRBs. Experiments with electron-positron pair plasmas will also become feasible in this next decade.

*Magnetic fields* in laboratory plasmas are proving to be larger and more complex than anticipated (Rygg et al., 2008), as may indeed also be true in stars. The combination of experiments and 3D modeling will produce large advances in understanding such fields over the next decade. In addition, understanding these effect of strong magnetic fields on atomic structure is essential to the analysis of spectra from the vicinity of magnetized compact objects, including neutron stars and black holes. Ggauss fields, large enough to have significant effects, are produced now (Wagner et al. 2004); the next decade will see 10 Ggauss.

*Dynamos* in stars are also poorly understood. Experiments are now testing some dynamo models in electrically conducting fluids (Monchaux et al. 2007), in turbulent plasmas (Ji et al. 1994), and in liquid metal (Spence et al. 2007). In the next decade, further progress is expected in elucidating mechanisms in generating magnetic field in MHD media and also in plasmas, and thus provide much needed understanding on the origin of magnetic field in stars and galaxies.

*Magnetic reconnection* occurs in solar flares and forming stars at rates orders of magnitude faster than classical theories can explain. Only recently have proposed faster mechanisms begun to be tested experimentally. For example, the conjectured Hall effect was verified (Ren et al., 2005). In the next decade, further progress is expected in the areas of reconnection in three dimensions, particle acceleration during reconnection, as well as reconnection rate scaling on system size.

What is required to realize the potential contributions in this area to astrophysics is for the relevant agencies to treat laboratory astrophysics as an important component of national research in plasmas, including but not limited to magnetized and high-energy-density plasmas.

### 3.5. Nuclear Physics

*Stellar Modeling and Neutrino Physics:* A decades-long effort to measure the reactions of the pp-chain and CN-cycle led to a calibrated standard solar model predicting patterns of neutrino fluxes inconsistent with experiment. This in turn motivated experiments that established that 2/3rds of the solar flux was in heavy-flavor neutrinos. More precise constraints on both the nuclear S-factors and the flavor physics will be coming from experiments. Newly precise solar neutrino experiments such as SNO+ have the potential (by measuring the CN-neutrino component of the solar flux) to determine directly the metalicity of the solar core. This is important because current solar metalicities that are inconsistent with interior solar sound speeds.

*Explosive astrophysics:* Approximately half of the elements heavier than iron were synthesized in core-collapse SN, by rapid neutron capture. One of the main goals of the planned Facility for Rare Ion Beams (FRIB) will be to produce the rare isotopes involved in the r-process for the first time, so that their masses and beta decay lifetimes can be measured. This will enable the nucleosynthetic output of SNe to be used to constrain aspects of the explosion, such as the dynamic timescale for the ejecta and the freezeout radius. These studies will also provide clues to the nature of the first stars and earliest nucleosynthesis in the Galaxy.

*Supernova physics:* Laboratory nuclear astrophysics provides critical input to SN modeling, including the Gamow-Teller strength distributions that govern neutrino-matter interactions and nucleosynthesis. The results will be incorporated into multi-D SN models now under development. By detecting and interpreting neutrinos from the next galactic SN, one may be able to constrain the unknown third mixing angle $\theta_{13}$, and to isolate a new aspect of the MSW mechanism, oscillations altered by an intense neutrino background.

*Neutron stars:* The structure of neutron stars impacts problems as diverse as the phases of QCD at high density (e.g., color-flavor locking) and the gravitational wave forms needed for the future data analysis. A key uncertainty is the precise value of the nuclear symmetry energy, which is related to the pressure of a neutron gas. A major experimental program at the Jefferson Laboratory to determine the neutron distribution of a heavy nucleus will soon address this.

*Nuclear astrophysics in heavier stars:* There are major uncertainties in our understanding of the nuclear physics governing massive star evolution. The most critical parameter in the hydrostatic evolution is the $^{12}C(\alpha,\gamma)$ cross section, which determines the C to O mass ration, and thus influences nucleosynthesis. There are several experimental efforts underway to further constrain this cross section, which is complicated due to the contributions of subthreshold resonances.

*Transient thermonuclear reactions:* Thermonuclear explosions following accretion on neutron stars are responsible for the X-ray bursts studied by observatories such as Beppo-SAX, Chandra, XMM-Newton, RXTE, and INTEGRAL. The details of these bursts are poorly understood. The observables depend on the nuclear physics of neutron-deficient nuclei participating in the $\alpha p$- and rp processes.FRIB will make significant progress in this area, specifically by the reacceleration of beams produced by fragmentation and gas stopping.

### 3.6 Particle Physics

*New, very light particles* are predicted by certain theories in particle physics. For example, the axion appears in models addressing the so-called strong CP problem. Such axions might be produced in stars and could dramatically affect their lifetime in the red giant phase. Many key constraints on axion properties come from stellar evolution. Several laboratory experiments will search for axions by looking for the conversion to photons in strong magnetic fields.

## 4. Four central questions and one area with unusual discovery potential

### 4.1 Four central questions
- What is the detailed, time-dependent structure of stars, and how does this affect their evolution and the consequences of slow neutron capture?
- What are the explosion mechanisms of SNe and what is the impact of these on nucleosynthesis?
- What are the mechanisms of mass loss from intermediate mass stars and how does such mass loss impact the structure and chemical evolution of the ISM ?
- What is the actual behavior of energetic events such as GRBs and how do these impact cosmic-ray acceleration?

## 4.2 One area with unusual discovery potential

- The role of magnetic fields in stars and stellar evolution. Because they are difficult to measure, even outside stars, and difficult to model, which requires three dimensions for any realism, laboratory studies have the potential to make seminal discoveries.

## 5. Postlude

Laboratory astrophysics and complementary theoretical calculations are the part of the foundation for our understanding of SSE and will remain so for many generations to come. From the level of scientific conception to that of the scientific return, it is our understanding of the underlying processes that allows us to address fundamental questions in these areas. In this regard, laboratory astrophysics is much like detector and instrument development; these efforts are necessary to maximize the scientific return from astronomical observations.

## References

Akiyama, S., et al. 2002, ApJ., 584, 954.
Badenes, C., et al. 2008, ApJ, 680, L33.
Bailey, J. E., et al. 2009, Physics of Plasmas in press
Basu, S. and H.M. Antia 2008, Physics Reports 457, 217.
Bryans, P. et al. 2009, ApJ, 691, 1540.
Drake et al. 2004, Phys. Plasmas 11, 2829.
Foord, M. E., et al. 2006, JQSRT 99, 712.
Ji, H., et al. 1994, Phys. Rev. Lett. 73, 668
Kuranz, C.C., et al. 2009, ApJ, in press.
Laming, J. M. 2009, arXiv:0901.3350
Laming, J. M., and Hwang, U. 2004, ApJ, 597, 347.
Monchaux, R., et al. 2007, Phys. Rev. Lett. 98, 044502
Ren, Y. et al. 2005, Phys. Rev. Lett. 95, 055003
Rho, J. et al. 2008, ApJ, 673, 271
Rygg, J. R., et al. 2008, *Science* 319 1223.
Soderberg, A. M., et al. 2008, *Nature*, 453, 469.
Spence, E., et al. 2007, Phys. Rev. Lett. 98, 164503
Wagner, U., et al. 2004, *Phys. Rev. E*, 70(2)